\documentclass[twocolumn,showpacs,preprintnumbers,amsmath,amssymb,superscriptaddress]{revtex4-1}

\usepackage{graphicx}
\usepackage{dcolumn}
\usepackage{bm}
\usepackage[figuresright]{rotating}
\usepackage{fancyhdr}
\usepackage{enumerate}
\usepackage{indentfirst}
\usepackage{xspace}
\usepackage{color}

\begin{document}

\title{\bf Measurement of Inclusive Neutral Current $\pi^{0}$ Production on Carbon in a Few-GeV Neutrino Beam}

\affiliation{Institut de Fisica d'Altes Energies, Universitat Autonoma de Barcelona, E-08193 Bellaterra (Barcelona), Spain}
\affiliation{Department of Physics, University of Colorado, Boulder, Colorado 80309, USA}
\affiliation{Department of Physics, Columbia University, New York, NY 10027, USA}
\affiliation{Fermi National Accelerator Laboratory; Batavia, IL 60510, USA}
\affiliation{High Energy Accelerator Research Organization (KEK), Tsukuba, Ibaraki 305-0801, Japan}
\affiliation{Department of Physics, Imperial College London, London SW7 2AZ, UK}
\affiliation{Department of Physics, Indiana University, Bloomington, IN 47405, USA}
\affiliation{Kamioka Observatory, Institute for Cosmic Ray Research, University of Tokyo, Gifu 506-1205, Japan}
\affiliation{Research Center for Cosmic Neutrinos, Institute for Cosmic Ray Research, University of Tokyo, Kashiwa, Chiba 277-8582, Japan}
\affiliation{Department of Physics, Kyoto University, Kyoto 606-8502, Japan}
\affiliation{Los Alamos National Laboratory; Los Alamos, NM 87545, USA}
\affiliation{Department of Physics and Astronomy, Louisiana State University, Baton Rouge, LA 70803, USA}
\affiliation{Department of Physics, Massachusetts Institute of Technology, Cambridge, MA 02139, USA}
\affiliation{Department of Chemistry and Physics, Purdue University Calumet, Hammond, IN 46323, USA}
\affiliation{Universit$\grave{a}$ di Roma La Sapienza, Dipartimento di Fisica and INFN, I-00185 Rome, Italy}
\affiliation{Physics Department, Saint Mary's University of Minnesota, Winona, MN 55987, USA}
\affiliation{Department of Physics, Tokyo Institute of Technology, Tokyo 152-8551, Japan}
\affiliation{Instituto de Fisica Corpuscular, Universidad de Valencia and CSIC, E-46071 Valencia, Spain}

\author{Y.~Kurimoto}\affiliation{Department of Physics, Kyoto University, Kyoto 606-8502, Japan}
\author{J.~L.~Alcaraz-Aunion}\affiliation{Institut de Fisica d'Altes Energies, Universitat Autonoma de Barcelona, E-08193 Bellaterra (Barcelona), Spain}
\author{S.~J.~Brice}\affiliation{Fermi National Accelerator Laboratory; Batavia, IL 60510, USA}
\author{L.~Bugel}\affiliation{Department of Physics, Massachusetts Institute of Technology, Cambridge, MA 02139, USA}
\author{J.~Catala-Perez}\affiliation{Instituto de Fisica Corpuscular, Universidad de Valencia and CSIC, E-46071 Valencia, Spain}
\author{G.~Cheng}\affiliation{Department of Physics, Columbia University, New York, NY 10027, USA}
\author{J.~M.~Conrad}\affiliation{Department of Physics, Massachusetts Institute of Technology, Cambridge, MA 02139, USA}
\author{Z.~Djurcic}\affiliation{Department of Physics, Columbia University, New York, NY 10027, USA}
\author{U.~Dore}\affiliation{Universit$\grave{a}$ di Roma La Sapienza, Dipartimento di Fisica and INFN, I-00185 Rome, Italy}
\author{D.~A.~Finley}\affiliation{Fermi National Accelerator Laboratory; Batavia, IL 60510, USA}
\author{A.~J.~Franke}\affiliation{Department of Physics, Columbia University, New York, NY 10027, USA}
\author{C.~Giganti}
\altaffiliation[Present address: ]{DSM/Irfu/SPP, CEA Saclay, F-91191 Gif-sur-Yvette, France}
\affiliation{Universit$\grave{a}$ di Roma La Sapienza, Dipartimento di Fisica and INFN, I-00185 Rome, Italy}
\author{J.~J.~Gomez-Cadenas}\affiliation{Instituto de Fisica Corpuscular, Universidad de Valencia and CSIC, E-46071 Valencia, Spain}
\author{P.~Guzowski}\affiliation{Department of Physics, Imperial College London, London SW7 2AZ, UK}
\author{A.~Hanson}\affiliation{Department of Physics, Indiana University, Bloomington, IN 47405, USA}
\author{Y.~Hayato}\affiliation{Kamioka Observatory, Institute for Cosmic Ray Research, University of Tokyo, Gifu 506-1205, Japan}
\author{K.~Hiraide}
\altaffiliation[Present address: ]{Kamioka Observatory, Institute for Cosmic Ray Research, University of Tokyo, Gifu 506-1205, Japan}
\affiliation{Department of Physics, Kyoto University, Kyoto 606-8502, Japan}
\author{G.~Jover-Manas}\affiliation{Institut de Fisica d'Altes Energies, Universitat Autonoma de Barcelona, E-08193 Bellaterra (Barcelona), Spain}
\author{G.~Karagiorgi}\affiliation{Department of Physics, Massachusetts Institute of Technology, Cambridge, MA 02139, USA}
\author{T.~Katori}\affiliation{Department of Physics, Indiana University, Bloomington, IN 47405, USA}
\author{Y.~K.~Kobayashi}\affiliation{Department of Physics, Tokyo Institute of Technology, Tokyo 152-8551, Japan}
\author{T.~Kobilarcik}\affiliation{Fermi National Accelerator Laboratory; Batavia, IL 60510, USA}
\author{H.~Kubo}\affiliation{Department of Physics, Kyoto University, Kyoto 606-8502, Japan}
\author{W.~C.~Louis}\affiliation{Los Alamos National Laboratory; Los Alamos, NM 87545, USA}
\author{P.~F.~Loverre}\affiliation{Universit$\grave{a}$ di Roma La Sapienza, Dipartimento di Fisica and INFN, I-00185 Rome, Italy}
\author{L.~Ludovici}\affiliation{Universit$\grave{a}$ di Roma La Sapienza, Dipartimento di Fisica and INFN, I-00185 Rome, Italy}
\author{K.~B.~M.~Mahn}
\altaffiliation[Present address: ]{TRIUMF, Vancouver, British Columbia, V6T 2A3, Canada}
\affiliation{Department of Physics, Columbia University, New York, NY 10027, USA}
\author{C.~Mariani}\affiliation{Department of Physics, Columbia University, New York, NY 10027, USA}
\author{S.~Masuike}\affiliation{Department of Physics, Tokyo Institute of Technology, Tokyo 152-8551, Japan}
\author{K.~Matsuoka}\affiliation{Department of Physics, Kyoto University, Kyoto 606-8502, Japan}
\author{V.~T.~McGary}\affiliation{Department of Physics, Massachusetts Institute of Technology, Cambridge, MA 02139, USA}
\author{W.~Metcalf}\affiliation{Department of Physics and Astronomy, Louisiana State University, Baton Rouge, LA 70803, USA}
\author{G.~B.~Mills}\affiliation{Los Alamos National Laboratory; Los Alamos, NM 87545, USA}
\author{G.~Mitsuka}
\altaffiliation[Present address: ]{Solar-Terrestrial Environment Laboratory, Nagoya University, Furo-cho, Chikusa-ku, Nagoya, Japan}
\affiliation{Research Center for Cosmic Neutrinos, Institute for Cosmic Ray Research, University of Tokyo, Kashiwa, Chiba 277-8582, Japan}
\author{Y.~Miyachi}
\altaffiliation[Present address: ]{Department of Physics, Yamagata University, Yamagata, 990-8560 Japan}
\affiliation{Department of Physics, Tokyo Institute of Technology, Tokyo 152-8551, Japan}
\author{S.~Mizugashira}\affiliation{Department of Physics, Tokyo Institute of Technology, Tokyo 152-8551, Japan}
\author{C.~D.~Moore}\affiliation{Fermi National Accelerator Laboratory; Batavia, IL 60510, USA}
\author{Y.~Nakajima}\affiliation{Department of Physics, Kyoto University, Kyoto 606-8502, Japan}
\author{T.~Nakaya}\affiliation{Department of Physics, Kyoto University, Kyoto 606-8502, Japan}
\author{R.~Napora}\affiliation{Department of Chemistry and Physics, Purdue University Calumet, Hammond, IN 46323, USA}
\author{P.~Nienaber}\affiliation{Physics Department, Saint Mary's University of Minnesota, Winona, MN 55987, USA}
\author{D.~Orme}\affiliation{Department of Physics, Kyoto University, Kyoto 606-8502, Japan}
\author{M.~Otani}\affiliation{Department of Physics, Kyoto University, Kyoto 606-8502, Japan}
\author{A.~D.~Russell}\affiliation{Fermi National Accelerator Laboratory; Batavia, IL 60510, USA}
\author{F.~Sanchez}\affiliation{Institut de Fisica d'Altes Energies, Universitat Autonoma de Barcelona, E-08193 Bellaterra (Barcelona), Spain}
\author{M.~H.~Shaevitz}\affiliation{Department of Physics, Columbia University, New York, NY 10027, USA}
\author{T.-A.~Shibata}\affiliation{Department of Physics, Tokyo Institute of Technology, Tokyo 152-8551, Japan}
\author{M.~Sorel}\affiliation{Instituto de Fisica Corpuscular, Universidad de Valencia and CSIC, E-46071 Valencia, Spain}
\author{R.~J.~Stefanski}\affiliation{Fermi National Accelerator Laboratory; Batavia, IL 60510, USA}
\author{H.~Takei}
\altaffiliation[Present address: ]{Kitasato University, Tokyo, 108-8641 Japan}
\affiliation{Department of Physics, Tokyo Institute of Technology, Tokyo 152-8551, Japan}
\author{H.-K.~Tanaka}\affiliation{Department of Physics, Massachusetts Institute of Technology, Cambridge, MA 02139, USA}
\author{M.~Tanaka}\affiliation{High Energy Accelerator Research Organization (KEK), Tsukuba, Ibaraki 305-0801, Japan}
\author{R.~Tayloe}\affiliation{Department of Physics, Indiana University, Bloomington, IN 47405, USA}
\author{I.~J.~Taylor}
\altaffiliation[Present address: ]{Department of Physics and Astronomy, State University of New York, Stony Brook, NY 11794-3800, USA}
\affiliation{Department of Physics, Imperial College London, London SW7 2AZ, UK}
\author{R.~J.~Tesarek}\affiliation{Fermi National Accelerator Laboratory; Batavia, IL 60510, USA}
\author{Y.~Uchida}\affiliation{Department of Physics, Imperial College London, London SW7 2AZ, UK}
\author{R.~Van~de~Water}\affiliation{Los Alamos National Laboratory; Los Alamos, NM 87545, USA}
\author{J.~J.~Walding}\affiliation{Department of Physics, Imperial College London, London SW7 2AZ, UK}
\author{M.~O.~Wascko}\affiliation{Department of Physics, Imperial College London, London SW7 2AZ, UK}
\author{H.~B.~White}\affiliation{Fermi National Accelerator Laboratory; Batavia, IL 60510, USA}
\author{M.~J.~Wilking}
\altaffiliation[Present address: ]{TRIUMF, Vancouver, British Columbia, V6T 2A3, Canada}
\affiliation{Department of Physics, University of Colorado, Boulder, Colorado 80309, USA}
\author{M.~Yokoyama}\affiliation{Department of Physics, Kyoto University, Kyoto 606-8502, Japan}
\author{G.~P.~Zeller}\affiliation{Los Alamos National Laboratory; Los Alamos, NM 87545, USA}
\author{E.~D.~Zimmerman}\affiliation{Department of Physics, University of Colorado, Boulder, Colorado 80309, USA}

\collaboration{The SciBooNE Collaboration}\noaffiliation

\begin{abstract}
The SciBooNE Collaboration reports inclusive neutral current neutral
pion production by a muon neutrino beam on a polystyrene target ($\rm
C_{8}H_{8}$). We obtain $(7.7 \pm 0.5({\rm stat.}) \pm 0.5({\rm
sys.})) \times 10^{-2}$ as the ratio of the neutral current neutral
pion production to total charged current cross section; the mean
energy of neutrinos producing detected neutral pions is 1.1 GeV. The
result agrees with the Rein-Sehgal model implemented in our neutrino
interaction simulation program with nuclear effects.  The spectrum
shape of the $\pi^0$ momentum and angle agree with the model. We also
measure the ratio of the neutral current coherent pion production to
total charged current cross section to be $(0.7 \pm 0.4) \times
10^{-2}$.
\end{abstract}
\date{\today}
\pacs{13.15.+g, 13.60.Le, 25.30.Pt, 95.55.Vj}

\maketitle
\newcommand{\pzero}{$\pi^{\rm{0}}$}

\section{Introduction}
\label{sec:introduction}

Neutrino-nucleus cross sections have been intensively studied for
decades.  However, the precision and understanding of the cross
sections around 1 GeV are still not satisfactory.  The next generation
of neutrino oscillation experiments will search for sub-leading flavor
oscillation and charge-parity symmetry violation; the precision
needed for these searches drives the need for more accurate independent
measurements of neutrino cross sections
\cite{Itow:2002rk,Harris:2004iq}.  Although several interaction
channels contribute to the total neutrino-nucleus cross section in the
neutrino energy range of a few GeV, an understanding of neutral
current neutral pion production (NC\pzero) is especially important.
NC\pzero\ events form the largest ${\nu}_{\mu}$-induced background to
neutrino experiments measuring ${\nu}_{\mu} \rightarrow {\nu}_{e}$
oscillations in the neutrino energy range of a few GeV or less, such
as the T2K experiment \cite{Itow:2002rk}. NC\pzero\ events can mimic
$\nu_e$ signal events when, for example, one of the two photons
associated with ${\pi}^{\rm 0} {\rightarrow} {\gamma}{\gamma}$ is not
detected.

NC\pzero production has been measured by several past
experiments~\cite{Barish:1974fe,Derrick:1980xw,Krenz:1977sw,Lee:1976wr,Nienaber:1988js}. However,
their results have large uncertainty due to low statistics and are not
useful expressions for predicting electron backgrounds in
${\nu}_{\mu}\rightarrow{\nu}_{e}$ oscillation searches, since they are
typically given as ratios to the charged current (CC) single pion
production cross section, which is also poorly known. T2K
\cite{Itow:2002rk} uses a neutrino beam whose mean neutrino energy is
approximately 0.8 GeV. The experiment requires less than a 10\%
uncertainty on NC\pzero\ production to maintain high sensitivity for
the ${\nu}_{\mu} \rightarrow {\nu}_{e}$ oscillation search.  Recently,
two experiments published NC \pzero\ production results.  The K2K
collaboration reported NC \pzero\ production in water with a 1.3~GeV
mean neutrino energy beam\cite{Nakayama:2004dp}, finding their
measurement consistent with the Monte Carlo (MC) prediction based on
the Rein and Sehgal model \cite{Rein:1980wg}.  MiniBooNE reported the
yield and spectral shape of \pzero\,s as a function of \pzero\
momentum and the observation of NC coherent \pzero\ production in
mineral oil (${\rm C H_{2}}$) in neutrino beam of mean neutrino energy
0.7~GeV \cite{AguilarArevalo:2008xs}.  
The total NC\pzero\ cross section below 1 GeV has still not
been precisely measured yet.

Pions are produced mainly through two distinct mechanisms by neutrinos
with energies around 1 GeV. In the dominant mode, resonant pion
production, the neutrino interacts with a nucleon in the nucleus and
excites it to a baryonic resonance, such as $\Delta$ (1232), which
subsequently decays to a pion and a nucleon. The other mode, coherent
pion production occurs when the neutrino interacts with the target
nucleus so that no nuclear breakup occurs. Resonance production and
decay in a nuclear target differs from the case of the free nucleon
target. This is due to nuclear effects such as Fermi motion, Pauli
blocking, and the nuclear potential. In addition, produced mesons and
baryons interact with nuclear matter until they escape from the target
nucleus. Due to this final state interaction, the number, momenta,
directions and charge states of produced particles can be changed in
nuclear matter. Although there are several theoretical models of these
processes, their uncertainties are still large. To understand the
production mechanism and the nuclear effects, measurements of emitted
\pzero\ kinematics are very important.

Recent measurements of coherent pion production 
have drawn much attention. For CC coherent pion production, 
the K2K experiment placed a limit on the ratio of the CC 
coherent pion production to the total CC cross sections at 1.3~GeV\cite{Hasegawa:2005td}. 
This result was confirmed by the SciBooNE experiment \cite{Hiraide:2008eu}, 
although recent data from the SciBooNE collaboration suggest CC
coherent pion production at a level below the published limit
in both neutrinos and antineutrinos~\cite{Hiraide:2009ag,Tanaka:2009}.
Moreover,  evidence for NC coherent pion production
with neutrino energy less than 2 GeV has been reported by 
the MiniBooNE Collaboration \cite{AguilarArevalo:2008xs}. 
Hence, it is interesting to search for NC coherent \pzero\ production 
in the SciBooNE data.

In this paper, we present measurements of the NC\pzero\ interaction in
polystyrene ($\rm C_{8}H_{8}$) using the same neutrino beam as
MiniBooNE (with mean neutrino energy 0.7 GeV). We measure the ratio of
the total inclusive NC\pzero\ cross section to the total CC
cross section and kinematic distributions of the emitted {\pzero}s.
We also extract the fraction of coherent NC \pzero\ events from the
inclusive NC \pzero\ data sample. In these analyses, we define
NC\pzero\ events to be NC neutrino interactions with at least one
\pzero\ emitted in the final state from the target nucleus.

\section{Experiment Description}
\label{sec:experiment}

\subsection{Neutrino Beam}
\label{sec:beam}

The SciBooNE experiment detected neutrinos produced by the Fermilab
Booster Neutrino Beam (BNB). The same BNB beam is also serving the
MiniBooNE experiment. The BNB uses protons accelerated to 8 GeV
kinetic energy by the Fermilab Booster synchrotron.  Beam properties
are monitored on a spill-by-spill basis, and at various locations
along the BNB line. Transverse and directional alignment of the beam,
beam width and angular divergence, beam intensity and losses along the
BNB, are measured and used to monitor data quality. Protons strike a
71.1 cm long beryllium target, producing a secondary beam of hadrons,
mainly pions with a small fraction of kaons. A cylindrical horn
electromagnet made of aluminum surrounds the beryllium target to
sign-select and focus the secondary beam. 
For the
data set used in this measurement, the horn polarity was set to
neutrino mode, focusing (defocusing) particles with positive
(negative) electric charge. The neutrino beam is mostly produced in
the 50 m long decay region, mainly from $\pi^+\to\mu^+ \nu_{\mu}$
in-flight decays. See \cite{AguilarArevalo:2008yp} for further
details.

\subsection{Neutrino Detector}
\label{sec:detector}

The SciBooNE detector was located 100~m downstream from the beryllium
target on the axis of the beam.  The detector comprised three
sub-detectors: a fully active and finely segmented scintillator
tracker (SciBar), an electromagnetic calorimeter (EC), and a muon
range detector (MRD).  SciBar served as the primary neutrino target
for this analysis.

\begin{figure}[bp]
  \begin{center}
    \includegraphics*[scale=0.85]{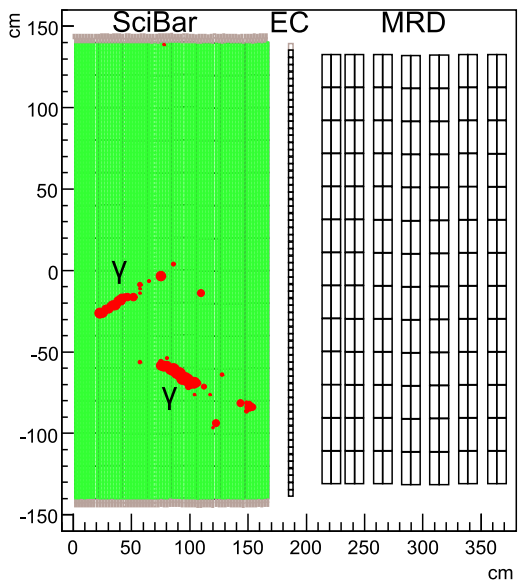}
    \caption{
    Event display of a typical NC\pzero\ event candidate in SciBooNE data.  
    The neutrino beam runs from left to right in this figure,
    encountering SciBar, the EC and MRD, in that order.
    The circles on SciBar indicate ADC hits for which the
    area of the circle is proportional to the energy deposition 
    in that channel. This event display shows the electromagnetic 
    shower tracks from the pair conversions of the two \pzero\ 
    decay photons. 
    }
      
    \label{fig:scibar_schematic}
  \end{center}
\end{figure}          

Fig.~\ref{fig:scibar_schematic} shows an event display of a typical
NC\pzero\ production event candidate.  SciBooNE
uses a right-handed Cartesian coordinate system in which the $z$ axis
is the beam direction and the $y$ axis is the vertical upward
direction.  The origin is located on the most upstream surface of
SciBar in the $z$ dimension, and at the center of the SciBar
scintillator plane in the $x$ and $y$ dimensions.  Since each
sub-detector is read out both vertically and horizontally, two views
are defined: top ($x\ vs.\ z$ projection) and side ($y\ vs.\ z$ projection).

The SciBar detector~\cite{Nitta:2004nt} was positioned upstream of the
other sub-detectors.  It consists of 14,336 extruded plastic
scintillator strips which serve as the target for the neutrino beam as
well as the active detection medium.  Each strip has a dimension of
1.3 $\times$ 2.5 $\times$ 300~cm$^3$.  The scintillators are arranged
vertically and horizontally to construct a 3 $\times$ 3 $\times$
1.7~m$^3$ volume with a total mass of 15 tons.  SciBar has about four
radiation lengths of material along the beam direction.  Each strip
was read out by a wavelength shifting (WLS) fiber attached to a
64-channel multi-anode PMT.  Charge information was recorded for each
channel, while timing information was recorded in groups of 32
channels by taking the logical OR with multi-hit TDC
modules~\cite{Yoshida:2004mh}.

The gains of all PMT channels, attenuation of WLS-fibers, and light
yield of each scintillator were continuously monitored \emph{ in situ}
using LED and cosmic ray data taken between beam spills, with precision
better than 1\%.  The timing resolution for minimum-ionizing
particles was evaluated with cosmic ray data to be 1.6~ns.  The
average light yield for minimum-ionizing particles is approximately 20
photoelectrons per 1.3~cm path length, and the typical pedestal width
is below 0.3 photoelectron.  The hit finding efficiency evaluated with
cosmic ray data is more than 99.8\%.  The minimum length of a
reconstructable track is approximately 8~cm (three layers hit in each
view).  The track finding efficiency for single tracks of 10~cm or
longer is more than 99\%.

The EC is located just downstream of SciBar, and is designed to
measure the electron neutrino contamination in the beam and tag
photons from $\pi^0$ decay.  The EC is a ``spaghetti'' type
calorimeter comprised of 1~mm diameter scintillating fibers embedded
in lead foil~\cite{Buontempo:1994yp}.  The calorimeter is made of 64
modules of dimensions 262 $\times$ 8 $\times$ 4~cm$^3$.  
The fibers are bundled in two independent groups of
4 $\times$ 4~cm$^2$ transverse cross section, read at both 
ends by Hamamatsu PMTs.
The EC comprises one
vertical and one horizontal plane (32 modules each), covering an
active area of 2.65 $\times$ 2.65~m$^2$.  The EC has a thickness of 11
radiation lengths along the beam direction.  The charge information
from each PMT was recorded.  A minimum ionizing particle with a
minimal path length deposits approximately 91~MeV in the EC.  The
energy resolution for electrons was measured to be 14\%$/\sqrt{E \
{\rm (GeV)}}$ using a test beam~\cite{Buontempo:1994yp}.  The
detection efficiency for cosmic muons is 96\%; the inefficiency
stems from gaps between the modules.

The MRD was installed downstream of the EC and is designed to measure
the momentum of muons produced by CC neutrino
interactions.  It had 12 iron plates with thickness 5~cm sandwiched
between planes of 6~mm thick scintillation counters; there were 13
alternating horizontal and vertical planes read out via 362 individual
2~inch PMTs.  Each iron plate measured 274 $\times$
305~cm$^2$.  The MRD measured the momentum of muons up to 1.2~GeV/$c$
using the observed muon range.  Charge and timing information from
each PMT were recorded.  Hit finding efficiency was continuously
monitored using cosmic ray data taken between beam spills; the average
hit finding efficiency is 99\%.

\subsection{Data Summary}
\label{sec:data}

The SciBooNE experiment took data from June~2007 until August~2008.
After applying data quality cuts to all beam
events\cite{Hiraide:2008eu}, $2.52\times 10^{20}$ protons on target are
usable for physics analysis, with $0.99\times 10^{20}$ protons on
target collected in neutrino mode.  The analysis presented herein uses
only neutrino mode data.

\section{Experiment Simulations}
\label{sec:simulation}

\subsection{Neutrino Flux Prediction}
\label{sec:flux}

Predictions for the BNB neutrino flux illuminating the SciBooNE
detector are obtained via a GEANT4 simulation of the beamline. The
simulation accounts for all relevant beamline geometry and materials,
the measured BNB beam optics properties, and the horn magnetic
field. Hadronic interactions are carefully modeled. Cross
sections for elastic, quasi-elastic and other inelastic interactions
of charged pions and nucleons with beryllium and aluminum are simulated
according to a custom model validated with external data, and covering
the most relevant momentum range (down to 0.5 GeV/c for pions, 2 GeV/c
for nucleons). Furthermore, the multiplicity and kinematics of all
relevant particle types produced in the inelastic interactions of
primary (8.4-8.9 GeV/c) protons with beryllium are also described by a
custom model based on external data. For $\pi^+$ production, a
parameterization based on HARP \cite{:2007gt} and BNL E910
\cite{:2007nb} data is used. Other hadronic and all electromagnetic
processes of importance to neutrino production are described instead
by standard GEANT4 models. The modeling of neutrino-producing weak
decays incorporates accurate knowledge of meson decay branching
fractions and form factors, and includes muon polarization
effects. For a detailed description of the BNB simulation code, see
\cite{AguilarArevalo:2008yp}.  According to the simulation, the
neutrino flux at the SciBooNE detector is dominated by muon neutrinos
(93\%), while the neutrino energy spectrum peaks at $\sim$ 0.6 GeV,
has a mean neutrino energy of $\sim$ 0.7 GeV, and extends up to 2-3
GeV \cite{Hiraide:2008eu}.

\subsection{Neutrino Interaction Simulation}
\label{sec:neutrinointeractionmc}

In the SciBooNE experiment, neutrino interactions with nuclear targets
are simulated by the NEUT program
library~\cite{Hayato:2002sd,Mitsuka:2008zz} that is used in the
Kamiokande, Super-Kamiokande, K2K, and T2K experiments.

The nuclear targets handled in NEUT are protons, carbon, oxygen, and
iron.  The energy of neutrinos handled by the simulation ranges from
100~MeV to 100~TeV.  The types of neutrino interactions simulated in
both NC and CC are : elastic and quasi-elastic
scattering ($\nu N \rightarrow \ell N'$), single meson production
($\nu N \rightarrow \ell N'm$), single gamma production ($\nu N
\rightarrow \ell N' \gamma$), coherent $\pi$ production ($\nu^{12}{\rm
C (or } ^{56}{\rm Fe}) \rightarrow \ell \pi\ ^{12}{\rm C (or }
^{56}{\rm Fe})$), and deep inelastic scattering ($\nu N \rightarrow
\ell N'hadrons$), where $N$ and $N'$ are the nucleons (proton or
neutron), $\ell$ is the lepton (electron, muon or neutrino), and $m$
is the meson. In nuclei, interactions of the mesons and hadrons with
the nuclear medium are simulated following the neutrino interactions.

\subsubsection{Single meson production via baryon resonances}
\label{sec:intra_nuclear}
\begin{figure}[bhtp]
  \includegraphics[width=8cm]{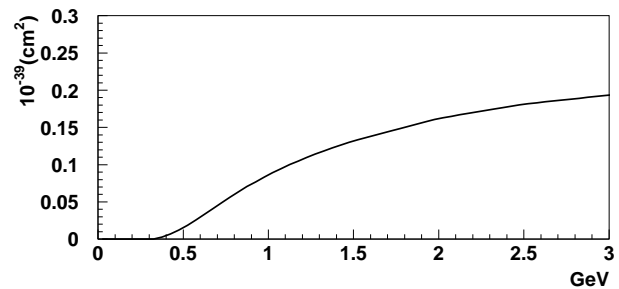}
  \caption{Cross section versus neutrino energy of NC single
$\pi^0$ production per nucleon on a polystyrene target ($C_8H_8$)
estimated in NEUT; the curve is based on the Rein and Sehgal model.
}
  \label{fig:nc1pi0}
\end{figure}

The main signal in this analysis is NC single $\pi^0$ production via
baryon resonances.  The resonant single meson production is simulated
based on the model of Rein and Sehgal~\cite{Rein:1980wg}.  The cross
section of the NC single $\pi^0$ production per nucleon on a
polystyrene target ($C_8H_8$) in NEUT is shown in
Fig.~\ref{fig:nc1pi0}.  The per nucleon cross section of a CH molecule
is calculated by summing the contributions from the six protons and
six neutrons bound in the carbon nucleus as well as the free proton,
and dividing that by 13.
Following production, the intra-nuclear interactions of the meson
and nucleons are simulated using a cascade model in which the
particles are traced until escaping from the nucleus. According to 
this model, approximately 40\% of {\pzero}s interact in the target
nucleus, averaged over our neutrino flux.
For scattering off nucleons in the nucleus by a neutrino, the
relativistic Fermi gas model of Smith and Moniz~\cite{Smith:1972xh} is
implemented. The nucleons are treated as quasi-free particles and the
Fermi motion of nucleons along with the Pauli exclusion principle is
taken into account.  The Fermi surface momentum is set to 217~MeV/$c$
and the nuclear potential is set to 27~MeV for carbon.  The vector and
axial-vector form factors are formalized to be dipole with
$M_V^{1\pi}=0.84$~$\rm GeV/c^2$ and $M_A^{1\pi}=1.21$~$\rm GeV/c^2$.  The same Fermi
momentum distribution, nuclear potential and $Q^2$-dependence of form
factors are used in all other neutrino-nucleus interactions except for coherent
$\pi$ production.

The Rein and Sehgal model assumes an intermediate baryon resonance,
$N^*$, in the reaction of $\nu N \rightarrow \ell N^*, N^* \rightarrow
N'm$.  All intermediate baryon resonances with mass less than
2~GeV/$c^2$ are included.  Baryon resonances with mass greater than
2~GeV/$c^2$ are simulated as deep inelastic scattering.  Pion-less
$\Delta$ decays---which produce no pion in the final state and account
for 20\% of $\Delta$ events~\cite{Singh:1998ha}---are also simulated.
To determine the angular distribution of final state pions,
Rein's method~\cite{Rein:1987cb} is used for the $P_{33}(1232)$
resonance. For other resonances, the directional distribution of the
generated pion is chosen to be isotropic in the resonance rest frame.

The inelastic scattering, charge exchange and absorption of pions in
nuclei are simulated. The interaction cross sections of pions in the
nuclei are based on the model by Salcedo {\it et
al}.~\cite{Salcedo:1987md}.  For inelastic scattering and charge
exchange interactions, the direction and momentum of pions are
affected.  In the scattering amplitude, Pauli blocking is also taken
into account.

\subsubsection{Coherent $\pi$ production}
\label{sssec:cohpisim}
\begin{figure}[bhtp]
  \includegraphics[width=8cm]{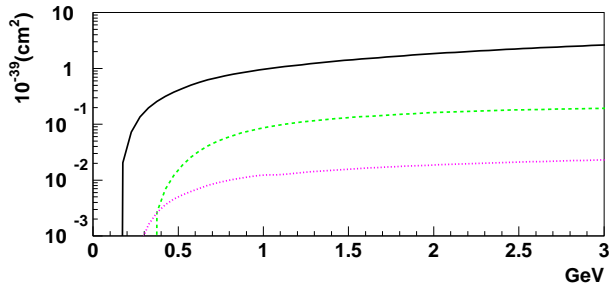}
  \caption{Cross sections versus neutrino energy of the total CC interaction (solid line), the NC single $\pi^0$ production via baryon resonances (dashed line) and the NC coherent $\pi^0$ production (dotted line) calculated in NEUT per nucleon on polystyrene target. }
  \label{fig:coherent}
\end{figure}

The \pzero\ signal events contain a contribution from NC coherent
$\pi^0$ production.  Due to the small momentum transfer to the target
nucleus, the outgoing neutrino and the pion tend to go in the forward
direction.  The formalism developed by Rein and
Sehgal~\cite{Rein:1982pf,Rein:2006di} is used to simulate the
interactions.  The axial vector mass, $M_A^{coherent}$, is set to 1.0~GeV/$c^2$,
and the nuclear radius parameter $R_0$ is set to 1.0~fm.  For the
total and inelastic pion-nucleon cross sections used in the formalism,
the fitted results given in Rein and Sehgal's paper are employed.  The
NC coherent $\pi^0$ production cross section on a polystyrene target
is shown in Fig.~\ref{fig:coherent}, with the NC single $\pi^0$
production via baryon resonances and the total CC cross
sections.  The Rein and Sehgal model predicts the NC coherent $\pi^0$
production rate to be approximately 1\% of the total neutrino CC
 rate in SciBooNE.

\subsubsection{Quasi-elastic scattering and deep inelastic scattering}

The dominant interaction in the SciBooNE neutrino energy is CC
quasi-elastic scattering, which is implemented using the Smith and
Moniz model~\cite{Smith:1972xh}. $M_V^{QE}$ and $M_A^{QE}$ are
set to be 0.84~$\rm GeV/c^2$ and 1.21~$\rm GeV/c^2$, respectively.

The deep inelastic scattering (DIS) cross section is calculated using
the GRV98 parton distribution functions~\cite{Gluck:1998xa}.  As well
as quasi-elastic scattering and deep inelastic scattering, other
neutrino interactions in NEUT are described in~\cite{Hiraide:2008eu}
in detail.

With the SciBooNE neutrino beam exposure of $0.99 \times 10^{20}$ protons on target, 
the expected number of events in the SciBooNE detector in each neutrino interaction is 
listed in Tab.~\ref{ta:NEUT}.
\begin{table}[tbp]
 \caption{The expected number and fraction of events in each neutrino interaction estimated by NEUT 
 at the SciBooNE detector location with the neutrino beam exposure of $0.99 \times 10^{20}$ protons on target.
The 10.6~ton fiducial volume of the SciBar detector is assumed.
CC and NC interactions are abbreviated as CC and NC, respectively.}
 \label{ta:NEUT}
 \begin{center}
  \begin{tabular}{lrr}
    \hline \hline
     Interaction Type                   &  \# Events & Fraction(\%) \\
    \hline
     CC quasi-elastic                   &  53,363  &  41.4  \\
     CC single $\pi$ via resonances     &  29,688  &  23.1  \\
     CC coherent $\pi$                  &   1,771  &   1.4  \\
     CC single meson except $\pi$       &     839  &   0.7  \\
     CC DIS                             &   6,074  &   4.7  \\
     NC elastic                         &  22,521  &  17.5  \\
     NC single $\pi^0$ via resonances   &   6,939  &   5.4  \\
     NC coherent $\pi^0$                &   1,109  &   0.9  \\
     NC single meson except $\pi^0$     &   4,716  &   3.7  \\
     NC DIS                             &   1,768  &   1.4  \\
    \hline \hline
  \end{tabular}
 \end{center}
\end{table}
For the purpose of systematic studies of neutrino interaction
simulations, we also use the NUANCE event
generator~\cite{Casper:2002sd} that is used in the MiniBooNE
experiment.  The types and models of neutrino interactions in NUANCE
are similar to those of NEUT but with different treatment of
re-interactions of mesons and hadrons with the nuclear medium.

\subsection{Neutrino detector simulation}
\label{subsec:detector_mc}

The GEANT4 framework is used for the detector simulation.  The Bertini
cascade model within GEANT4~\cite{Heikkinen:2003sc} is used to simulate
the interactions of hadronic particles with detector materials.
The detector simulation includes a detailed geometric model of the detector,
including the detector frame and experimental hall and soil, based on
survey measurements taken during detector construction.
A description of the detector simulation is given in~\cite{Hiraide:2008eu}.

In addition to neutrino interactions inside the detector, we also
simulate interactions in the surrounding material (the walls of the
detector hall and soil).  The density of material is assumed to be
2.15~$g/cm^3$ for the calculation of the interaction rate, and
concrete of that density is used as the material for propagation
of product particles.  We generate events in a volume
of $\pm$5~m in $x$, $y$, and $z$ direction in the SciBooNE coordinates.

\section{Data analysis}
\label{sec:analysis}
The present analysis has two main goals.   The first is to 
measure the ratio of NC\pzero\ production cross section to the total 
CC cross section. We measure the ratio of cross sections 
in order to minimize systematic uncertainty due to the neutrino flux prediction.
The second goal is to measure
the \pzero\ momentum and angular spectra.
In addition to the two main goals, we also extract 
the coherent \pzero\ fraction in the context of the Rein and Sehgal model. 

We reconstruct gamma rays converting in SciBar 
and select events with two reconstructed gamma rays and no 
muons, which is the characteristic topology of NC\pzero events.
We do not include NC\pzero events in which one or both gamma rays convert in the EC. 

\subsection{Signal Definition}
\label{subsec:sigdef}

We define an NC\pzero\ interaction as an NC neutrino interaction in
which at least one \pzero\ is emitted in the final state from the
target nucleus, $\nu_{\mu}C\rightarrow\nu_{\mu}\pi^0X$ where $X$
represents the nuclear remnant and any combination of nucleons and
mesons.  According to our MC simulation, 96\% of NC\pzero\ events
without any selection cuts have a single \pzero\ (85\% from a single
\pzero\ without any other mesons and 11\% from a single \pzero\ with
charged mesons) and 4\% have two {\pzero}s.  Any \pzero\ emitted from
the initial target nucleus constitutes a signal event whether it is
created from the neutrino vertex or final state interactions. Events
with a {\pzero} produced in the neutrino interaction but absorbed in
the target nucleus are not included in the signal sample, nor are
events in which {\pzero}s are produced by secondary particles
interacting with the detector scintillator outside the target nucleus.

\subsection{Gamma Ray Reconstruction}
\label{subsec:evrec}

\subsubsection{Gamma Conversion Probability}
Since the length of SciBar in the beam direction corresponds to four radiation lengths, 
a significant fraction of gamma rays escape from SciBar
without conversion. In 30\% of events with a \pzero\ emitted within SciBar's fiducial volume, 
both gamma rays convert in SciBar; in 38\%, only one 
gamma ray converts in SciBar; in 32\%, neither gamma ray
converts in SciBar. Since we aim to reconstruct two gamma rays
to identify NC\pzero\ events, the maximum detection efficiency 
attainable is 30\%.

\subsubsection{Track Reconstruction}
The first step of the event reconstruction is to search for
two-dimensional tracks in each view of SciBar using a cellular
automaton algorithm \cite{Maesaka:2005aj}.  For tracking, the hit
threshold is set to 2.5 photo-electrons, corresponding to roughly
0.25~MeV.  Three dimensional tracks are reconstructed by matching the
timing and $z$-edges of the two dimensional projections. In order to
match track projections in a three dimensional track, the timing
difference between two two-dimensional projections is required to be
less than 50 ns, and the $z$-edge difference must be less than 6.6~cm
for upstream and downstream edges. This method is
used for all charged particles and is the first step of gamma ray
reconstruction.

\subsubsection{Particle Identification Parameter}
The SciBar detector has the capability to 
distinguish protons from other particles using $dE/dx$
since recoil protons at SciBooNE energies interact well above 
minimum ionizing energy deposit.  We define a muon~confidence~level (MuCL)
using the observed energy deposit per layer for all reconstructed 
tracks\cite{Hiraide:2008eu}. The MuCL of the proton tracks 
tends to be close to 0 while the MuCL of other tracks tends
to be close to 1. Proton-like tracks are defined to have MuCL less than 0.03.

\subsubsection{Extended Track}
\label{subsubsec:etrack}
Single reconstructed tracks are extended in two ways to improve the energy 
reconstruction of gamma rays within SciBar. The first step is merging 
two or more tracks if they are nearly co-linear, because 
electromagnetic showers can form separate hit clusters in SciBar resulting
in two or more tracks. 
The second step is collecting lone hits around merged tracks. 
Electromagnetic showers sometimes deposit energy around the main track 
and these hits are missed by the track reconstruction algorithm.
Hits not assigned to any track within 20 cm from 2D projections of 
the merged track (\emph{i.e.}, after the first step) for each view are
added to the extended track.
The methods described above are applied only to non-proton-like tracks.
For the energy reconstruction, we use charge information 
of hits associated with original tracks as well as newly assigned hits.
For reconstructing the directions of gamma rays, 
we fit positions of hits in all original tracks in the extended
track with a straight line and do not use hits newly collected in the second step.

\subsubsection{SciBar-EC matching}
Gamma rays can escape SciBar to deposit energy in the EC.
After event reconstruction in SciBar, we search for EC clusters
aligned with tracks from SciBar. One EC cluster is defined as
a collection of neighboring EC hits. 
For an EC hit, the pulse heights of both side PMTs are required to be 
above threshold, which is set to 3 times the width of each pedestal---about 7~MeV.
The energy of an EC hit is the geometric average of the two PMTs.
The center of an EC cluster is defined as the energy-weighted average of hits in the cluster.
To match an EC cluster to an SciBar track, the EC cluster is required 
to be within 10 cm of the extrapolated two-dimensional projections of the 
SciBar track in each EC plane.
The energy of matched EC clusters is added to the corresponding 
extended tracks.

\subsubsection{Gamma Ray Reconstruction Performance}
We study the performance of the gamma ray reconstruction algorithms
using the MC simulation.
The angular resolution of gamma rays passing all selection cuts 
(see section \ref{subsec:evsec}) is estimated 
to be approximately 6$^{\circ}$.
For the energy reconstruction, 
energy of matched EC clusters is added to corresponding
extended tracks. About 7\% of selected 
extended tracks are made by two or more tracks.
The average gamma ray energy deposit of SciBar
is estimated to be 116 MeV and the energy resolution 
is estimated to be 6\%. About 17\% of selected 
extended tracks have matched EC clusters. The 
average gamma ray energy deposit in such matched EC clusters
is estimated to be 72 MeV and the energy resolution is estimated
to be 32\%.

Not all gamma ray energy is deposited in, nor recorded by, the detector.
Such lost energy is called leakage.
The actual leakage, defined as 
\begin{eqnarray}
L_{act} = 1 - \frac{\rm  gamma~energy~in~extended~track}{\rm true~gamma~energy},
\end{eqnarray}
is estimated to be 24\%; 11\% comes from energy loss in passive regions
and gamma rays escaping from the detectors 
and 13\% comes from energy deposit in active regions but 
not assigned to the extended track.
The reconstructed energy can include energy deposited by other particles.
On average, 15\% of the total energy in an extended track comes from other particles.
Due to this contamination, the effective leakage, defined as: 
\begin{eqnarray}
  L_{eff}=1 - \frac{\rm  reconstructed~energy~of~extended~track}{\rm true~gamma~energy},
\end{eqnarray}
is 15\%, which is smaller than $L_{act}$ averaged over all NC\pzero\ events.

\subsection{CC Event Selection}
\label{subsec:ccsec}
To identify CC events, we search for 
events in which at least one reconstructed track in
SciBar, when projected out of SciBar, is matched with a track or hits in the MRD. 
We reject events with hits associated with the muon track on
the most upstream layer of SciBar to eliminate neutrino-induced incoming
particles from the upstream wall or soil. 
The neutrino interaction vertex
for CC events is reconstructed as the upstream edge of the muon track.  
We select events whose vertices are in the SciBar fiducial
volume, defined to be $\pm$130~cm in both the $x$ and $y$ dimensions,
and 2.62~cm$<z<$157.2~cm, a fiducial mass of 10.6~tons. 
The time of the muon track is required to be within a 2~$\mu$s window
around the beam pulse. Finally, we require the muon track to stop in the MRD.
The MRD-stopped event sample~\cite{Hiraide:2008eu} serves as the normalization sample in the 
cross-section ratio measurement.
Unless otherwise indicated, the MC distributions in this paper are normalized 
using the MRD-stopped data sample.

\subsection{NC\pzero\ Event Selection}
\label{subsec:evsec}

The clearest feature of NC \pzero\ production is two gamma rays, 
coming from the decay of the \pzero, converted into two $e^+e^-$ pairs.
Background events stem from sources both internal and external to SciBar.
Internal backgrounds are neutrino interactions other than NC\pzero (mainly CC) within SciBar; 
external backgrounds come from neutrino interactions in the material 
outside of the detector volume (support structure, walls and soil---so called dirt events) 
as well as from cosmic rays.
In dirt events, neutrinos interact with materials such as the walls of
the  experimental hall or soil and produce secondary particles which 
deposit energy within SciBar's fiducial volume and cause recorded hits. 
The contribution of accidental cosmic rays in any event sample is small
 and accurately estimated by data taken with off-beam timing;
the fraction of accidental cosmic ray events 
is 1.8\% after all event selection cuts. Data distributions
shown hereafter have been cosmic-ray subtracted from them.
The event
selection cuts for NC\pzero\ production are developed
to select events with  two gamma rays while rejecting these backgrounds.

\subsubsection{Pre-selection Cuts}
\label{sssec:pre}
We select events with at least two three-dimensional tracks.
The timing of tracks are required to be within 50 ns of each other
and the closest endpoints of any two tracks are required to be 
in the fiducial volume defined in Sec.~\ref{subsec:evsec}.
The closest endpoints are chosen as the edge combination 
for two given tracks that gives the minimum distance among the four 
possible combinations. The times of the two tracks are required
to be within the 2~$\mu$s beam window.
In addition, we reject events with hits in the first layer of SciBar
within 100 ns from the times of tracks, to remove dirt background events.

\subsubsection{Muon Track Rejection Cuts}
\label{sssec:noside}
Events with muons are predominantly CC events.
To avoid muons which escape SciBar but do not penetrate the MRD,
we reject events with tracks escaping from the side of SciBar;
both edges of tracks must be in $\pm$130~cm in both
the $x$ and $y$ dimensions.

\begin{figure}[tbp]
  \begin{center}
    \includegraphics[keepaspectratio=false,height=70mm,width=70mm,trim=20mm 10mm 0mm 0mm]{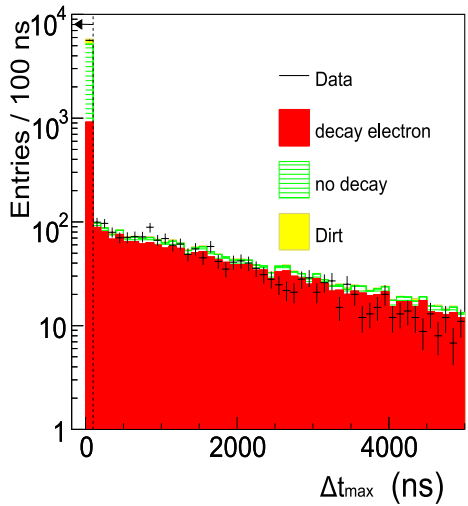}
  \end{center}
  \caption{${\Delta}t_{\rm max}$ distribution
(\ref{sssec:noside}) after rejection of side escaping tracks.  The
contributions from events with decay electrons, events without decay
electrons and the dirt events are shown separately for the MC
simulation.  }

  \label{fig:tdiff_max_nominal}
\end{figure}

To reject muons stopping in SciBar, we tag the electrons from muon decay.
No charge information is recorded for any scintillator strip after its 
first hit in an event, but the times of hits above threshold continue to be recorded.
Thus most decay electrons are not reconstructed as tracks but can be
identified as delayed time hits near the end of a muon track. 
We examine the  maximum timing difference (${\Delta}t_{\rm max}$) between the track times
and late hits times at the ends of tracks.
Events with decay electrons yield high values of ${\Delta}t_{\rm max}$
because of the long muon lifetime (${\tau}_{\mu}~=~2.2~{\mu}~{\rm sec}$).
Figure~\ref{fig:tdiff_max_nominal} shows the  ${\Delta}t_{\rm max}$ distribution.
We select events with ${\Delta}t_{\rm max}$ less than 100~ns.  

\subsubsection{Track Disconnection Cut}
 
\begin{figure}[tbp]
  \begin{center}
    \includegraphics[keepaspectratio=false,height=70mm,width=70mm,trim=0mm 10mm 0mm 0mm]{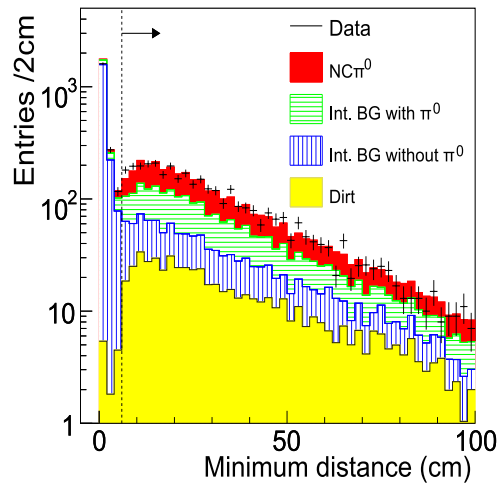}
  \end{center}
  \caption{The minimum distance between two tracks after
    muon rejection cuts.  The contribution from NC\pzero\ production,
    internal backgrounds with a {\pzero} in the final state, internal
    background without a {\pzero} in the final state and 
    ``dirt'' background events are shown separately for the MC
    simulation.}
  \label{fig:proj_distance_zoom_logy}
\end{figure}

CC events often have two or more tracks with a common vertex
while two gamma rays from {\pzero}s are typically isolated from each other. 
Hence, the distance between two tracks is used to separate two gamma rays from
CC events. 
If there are two particles with a common vertex, the minimum distance 
is close to zero. Figure~\ref{fig:proj_distance_zoom_logy} shows the distribution
of the minimum distance. 
Events with a minimum distance greater than 6 cm are selected.

\subsubsection{Electron Catcher Cut}
\label{sssec:eccut}

\begin{figure}[tbp]
 \begin{center}
   \includegraphics[height=70mm,width=70mm,trim=0mm 10mm 0mm 0mm]{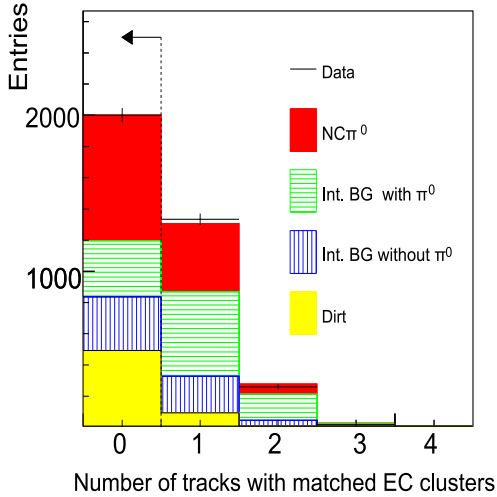}
  \end{center}
 \caption{Option 1 of the EC cut: the number of track with matched EC
clusters after the track disconnection cut. Events without SciBar-EC
matched tracks pass the EC cut.  Events with SciBar tracks matching EC
clusters can pass the EC cut if they satisfy the requirement shown in
Fig.~\ref{fig:edep_upstream_logy} or \ref{fig:edep_ratio_logy}}
 \label{fig:neccluster}
\end{figure}

\begin{figure}[tbp]
 \begin{center}
   \includegraphics[height=70mm,width=70mm,trim=0mm 10mm 0mm 0mm]{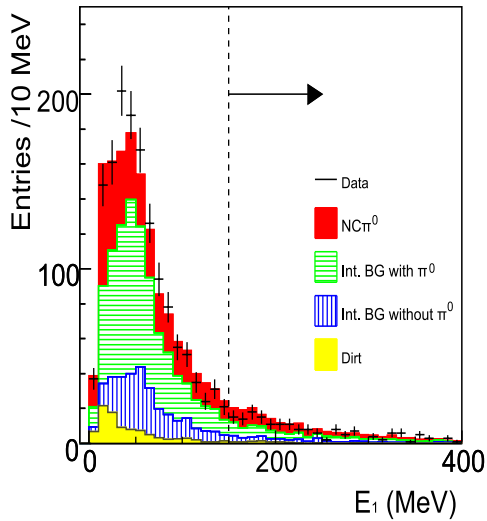}
  \end{center}
 \caption{Option 2 of the EC cut: the energy deposited in the upstream
layer of the EC ($E_{\rm 1}$) after the track disconnection
cut. Events with $E_{\rm 1}~>~150~{\rm MeV}$ pass the EC cut. Events
with $E_{\rm 1}~<~150~{\rm MeV}$ can pass the EC cut if they satisfy
the requirement shown in Fig.~\ref{fig:edep_ratio_logy}}
 \label{fig:edep_upstream_logy}
\end{figure}

\begin{figure}[tbp]
 \begin{center}
   \includegraphics[height=70mm,width=70mm,trim=0mm 20mm 0mm 0mm]{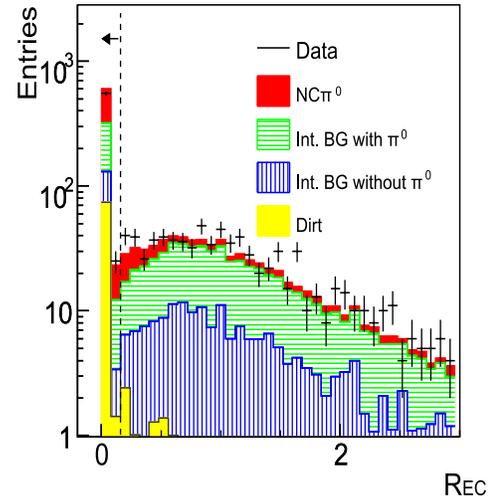}
  \end{center}
 \caption{Option 3 of the EC cut: the ratio of deposited energy in the
 downstream over the upstream layer (${\rm R}$) after the track
 disconnection cut. Only events with $E_{\rm 1}~<~150~{\rm MeV}$ are
 shown.}
 \label{fig:edep_ratio_logy}
\end{figure}

Matched EC Clusters are used to reject muons penetrating the EC.
Two quantities are used: the energy deposit in matched EC clusters 
in the upstream layer, $E_{\rm 1}$ ($E_{\rm 2}$ is energy in the downstream layer),
and the ratio of energy deposits in the downstream EC cluster over the upstream EC cluster,
${\rm R_{EC}}=E_{\rm 2}/E_{\rm 1}$.
Figure~\ref{fig:neccluster}, \ref{fig:edep_upstream_logy} and \ref{fig:edep_ratio_logy} shows 
the distributions of the number of tracks with matched EC clusters, $E_{\rm 1}$ and  ${\rm R_{EC}}$,
respectively.
Events satisfying one of the three following 
condition are selected, (i) No matched EC clusters or (ii) $E_{\rm 1}~>~150~{\rm MeV}$
or (iii) ${\rm R_{EC}}~<~0.2$.  
For events without matched EC clusters, both $E_{\rm 1}$
and ${\rm R_{EC}}$ are left undefined (not included in Fig.~\ref{fig:edep_upstream_logy}
and \ref{fig:edep_ratio_logy}).
For events with only upstream matched EC clusters, the minimum $E_{\rm 1}$
is chosen and ${\rm R_{EC}}$ is set to 0. For events with both upstream and
downstream matched EC clusters, the maximum ${\rm R_{EC}}$ is chosen
and the corresponding $E_{\rm 1}$ is chosen.

\subsubsection{Number of Photon Tracks}
\label{ssxec:extrk} 
\begin{figure}[tbp]
  \begin{center}
    \includegraphics[keepaspectratio=false,width=70mm,trim=0mm 10mm 0mm 0mm]{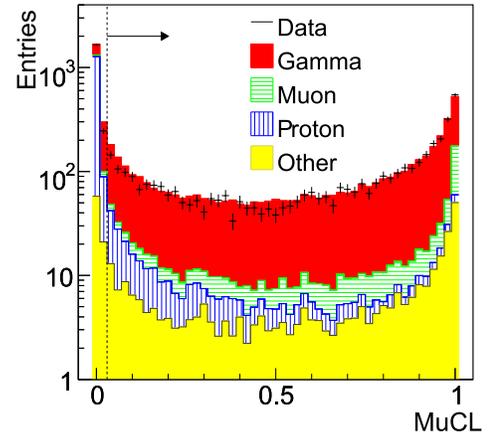}
  \end{center}
  \caption{The muon confidence level (MuCL) of tracks after the EC cut.
     The contributions from true gamma ray, muon, and proton tracks 
     are shown separately for the MC simulation.
   }
  \label{fig:mucl_long_particle_logy}
\end{figure}

We use the extended track information instead of the track 
information hereafter. As described in Sec.\ref{subsubsec:etrack},
we only use non-proton-like tracks to obtain extended tracks. 
Fig.~\ref{fig:mucl_long_particle_logy} shows the MuCL distribution after the EC cut. 
The dashed line (${\rm MuCL} = 0.03$) separates into particles to proton-like 
or non-proton-like tracks. The gamma ray efficiency for the non-proton-like sample
is 87\% and the contamination of gamma rays in the non-proton-like sample is 81\%.

\begin{figure}[tbp]
 \begin{center}
   \includegraphics[keepaspectratio=false,width=70mm,trim=0mm 10mm 0mm 0mm]{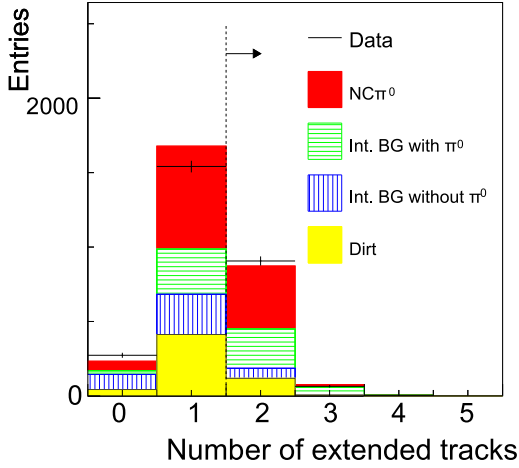}
  \end{center}
 \caption{The distribution of the number of extended tracks after the EC cut.}
 \label{fig:netracks}
\end{figure}

Figure~\ref{fig:netracks} shows the distribution
of the number of extended tracks after the EC cut. 
To reconstruct {\pzero}s, events with more than one extended track are selected. 
As shown in Fig.~\ref{fig:netracks}, 58\% of the NC\pzero\ events 
have only one extended track and are rejected by this cut. 
However, 39\% are events with one reconstructed gamma ray,
due to mis-reconstruction or gamma rays not converting in SciBar. 
Such events can not be used for \pzero\ reconstruction; 
12\% are events where the two gamma rays are reconstructed 
as two tracks but one of them is identified as a proton-like track; 
7\% are events in which the two gamma rays are reconstructed as two tracks 
but then merged, resulting in one extended track.

This cut is also effective at rejecting dirt backgrounds since 
many dirt background events have only one extended track, 
as shown in Fig.~\ref{fig:netracks}.

\subsubsection{Reconstructed {\pzero} Vertex Position Cut} 
\begin{figure}[tbp]
 \begin{center}
   \includegraphics[keepaspectratio=false,width=70mm,trim=0mm 10mm 0mm 0mm]{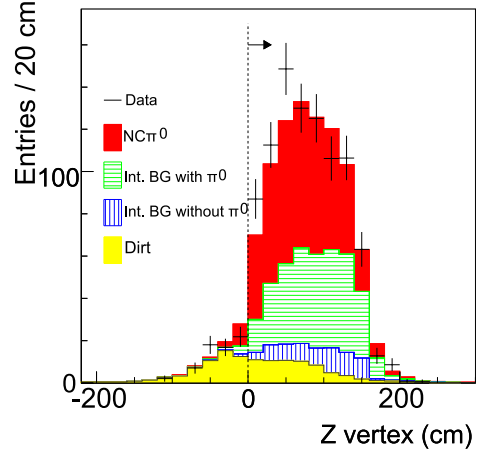}
  \end{center}
 \caption{The reconstructed $z$-vertices of {\pzero}s after the requirement of at least 
two extended tracks.}
 \label{fig:vtx_z_merged}
\end{figure}

The reconstructed vertex position of a \pzero\ is calculated as 
the intersection of two two-dimensional extended tracks. 
First, we calculate the intersection point for all combinations
of extended tracks in each view---two $z$ positions ($z_{\rm top}$
and $z_{\rm side}$) for each combination. 
We choose the combination giving the minimum  $|z_{\rm top}-z_{\rm side}|$ 
as the \pzero\ candidate. The reconstructed $z$-vertices are 
obtained by taking the error weighted average of $z_{\rm top}$
and $ z_{\rm side}$:
\begin{eqnarray}
z = \frac{\frac{z_{top}}{{{\delta}z_{top}}^{2}} + \frac{z_{side}}{{{\delta}z_{side}}^{2}}}{\frac{1}{{{\delta}z_{top}}^{2}} + \frac{1}{{{\delta}z_{side}}^{2}}},
\end{eqnarray}   
where ${\delta}z_{top(side)}$ is the error on $z_{\rm top(side)}$ returned by the track reconstruction algorithm. 
Figure~\ref{fig:vtx_z_merged} shows the reconstructed $z$-vertices of {\pzero}s. 
The vertex resolution is approximately 12 cm for all three dimensions.
Most events with a \pzero\ produced in SciBar yield a vertex within SciBar---but 
many dirt events yield a vertex position upstream of SciBar---so we select 
events with reconstructed \pzero\ $z$-vertex greater than 0~cm.

\subsubsection{Reconstructed \pzero\ Mass}
\label{sssec:pi0masscut} 
\begin{figure}[tbp]
 \begin{center}
   \includegraphics[keepaspectratio=false,width=70mm,trim=0mm 10mm 0mm 0mm]{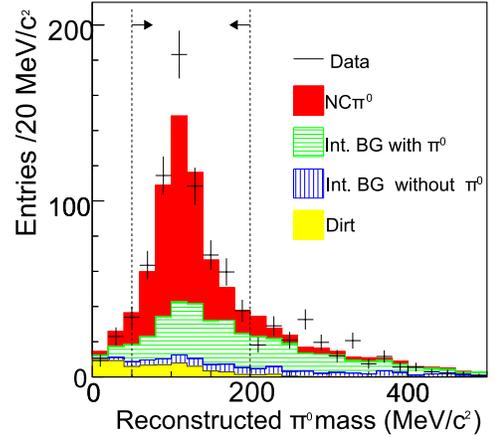}
  \end{center}
 \caption{The reconstructed mass of {\pzero}s after the reconstructed vertex position cut.}
 \label{fig:pi0mass_merged}
\end{figure}

Figure~\ref{fig:pi0mass_merged} shows the reconstructed mass of the \pzero\
calculated as  $\sqrt{2\rm{E^{rec}_{{\gamma}1}}\rm{E^{rec}_{{\gamma}2}}(1-\cos{{\theta}^{\rm{rec}}})}$, 
where $\rm{E^{rec}_{{\gamma}1}}$ and $\rm{E^{rec}_{{\gamma}2}}$ are the 
reconstructed energies of the extended tracks ($\rm{E^{rec}_{{\gamma}1}} > \rm{E^{rec}_{{\gamma}2}}$) 
and ${\theta}^{\rm{rec}}$ is the reconstructed angle between the extended tracks. 
The MC simulation describes well the tail of the distribution, which 
is background-dominated.
We select events with
 $ 50~{\rm MeV}/{c^{2}} < \rm{M^{rec}_{\pi^{\rm{0}}}} < 200~{\rm MeV}/{c^{2}} $.
The peak value is smaller than
 the actual \pzero\ mass (135 MeV) due to energy leakage of $\gamma$s.

\subsubsection{Event Selection Summary}
 \label{sssec:evsum}
Table~\ref{tab:evsum} shows the number of events in data and the MC at
each event selection stage. The numbers for the MC simulation 
are normalized to the number of MRD stopped events.
We select 657 events after all cuts. 
Subtracting the estimated background of 240 events (202 internal and 38 external) yields 
417 signal events.  The MC expectation is 368 events.
The purity of NC \pzero\ production after all event selection cuts is
estimated to be 61\% (40\% from single $\pi$ production via resonance decay,
 15\% from coherent $\pi$ production and 5\% from neutrino deep inelastic scattering).
According to our MC simulation, 96\% of selected NC\pzero\ events have one \pzero\ 
(91\% from a single \pzero\ without any other mesons
and 5\% from a single \pzero\ with charged mesons)
and 4\% have two {\pzero}s.
The efficiency for NC\pzero\ production, defined as: 
\begin{eqnarray}
\epsilon_{NC\pi^0} = \frac{{\rm the~number~of~selected~NC}{\pi}^{0}{\rm ~events}}{{\rm the~number~of~generated~NC}{\pi}^{0}{\rm ~events}},
\end{eqnarray}
is estimated to be 5.3\%. 
The internal background, which accounts for 33\% of this sample, contains 
CC \pzero\ production including secondary {\pzero}s (18\%), NC secondary \pzero\ production in detector materials (9\%) 
and non-\pzero\ background (6\%).  
According to our MC simulation, the average energy of neutrinos producing 
NC\pzero\ events in the SciBar fiducial volume is 1.3~GeV and the
average energy of neutrinos producing NC\pzero\ events that pass all
selection cuts is 1.1~GeV.  The average energy of neutrinos producing
NC{\pzero}s coherently is 1.1~GeV, while the average energy that pass
the selection cuts is 1.0~GeV.

\begin{table*}[tbp]
 \caption{Event selection summary; for the MC expectation, 
 NC\pzero signals, integral backgrounds (BG)
 and dirt backgrounds are shown separately. The number 
 of CC background events is shown in parentheses.}
 
 \label{tab:evsum}
 \begin{center}
  \begin{tabular}{lrrrrrrr}
    \hline \hline
    Event selection             & DATA   & \multicolumn{3}{c}{MC}  & NC\pzero\ & NC\pzero\ \\
                               &        & NC\pzero\ signal &  Internal BG (CC) & Dirt BG & Efficiency & Purity\\
    \hline
    Pre-selection Cuts               & 11,926 & 1,893  & 9,808 (9050) & 895 & 27.3\% & 15\%    \\
    Muon Track Rejection Cuts           &  5,609 & 1,377  & 3,785 (3326) & 606 & 19.8\% & 24\%\\
    Track Disconnection Cuts       &  3,614 & 1,314  & 1,706 (1306) & 595 & 18.9\% & 36\%\\
  
    Electron Catcher cut                      &  2791  &  1202  &  1088 (714) & 579 & 17.3\% & 42\% \\
    Number of Photon Tracks          &    973 &   443  &   389 (294) & 121 &  6.5\% & 46\% \\
    \pzero\ Reconstructed \pzero\ Position Cut           &    905 &   428  &   382 (288) &  65 &  6.2\% & 49\%  \\
    Reconstructed \pzero\ mass              &    657 &   368  &   202 (140) &  38 &  5.3\% & 61\%  \\    
    \hline \hline
  \end{tabular}
 \end{center}
\end{table*}
\section{Results}
\label{sec:results}


\subsection{$\sigma(\rm{NC}\pi^{\rm{0}})/\sigma(\rm{CC})$ Cross Secion Ratio}
\label{chap:crosssection}
We measure the ratio of the NC$\pi^{\rm{0}}$ production to
the total CC interaction cross sections.

\subsubsection{NC$\pi^{\rm{0}}$ Production}
\begin{figure}[htbp]
  \begin{center}
      \includegraphics[keepaspectratio=false,width=70mm,trim=0mm 30mm 50mm 50mm,clip]{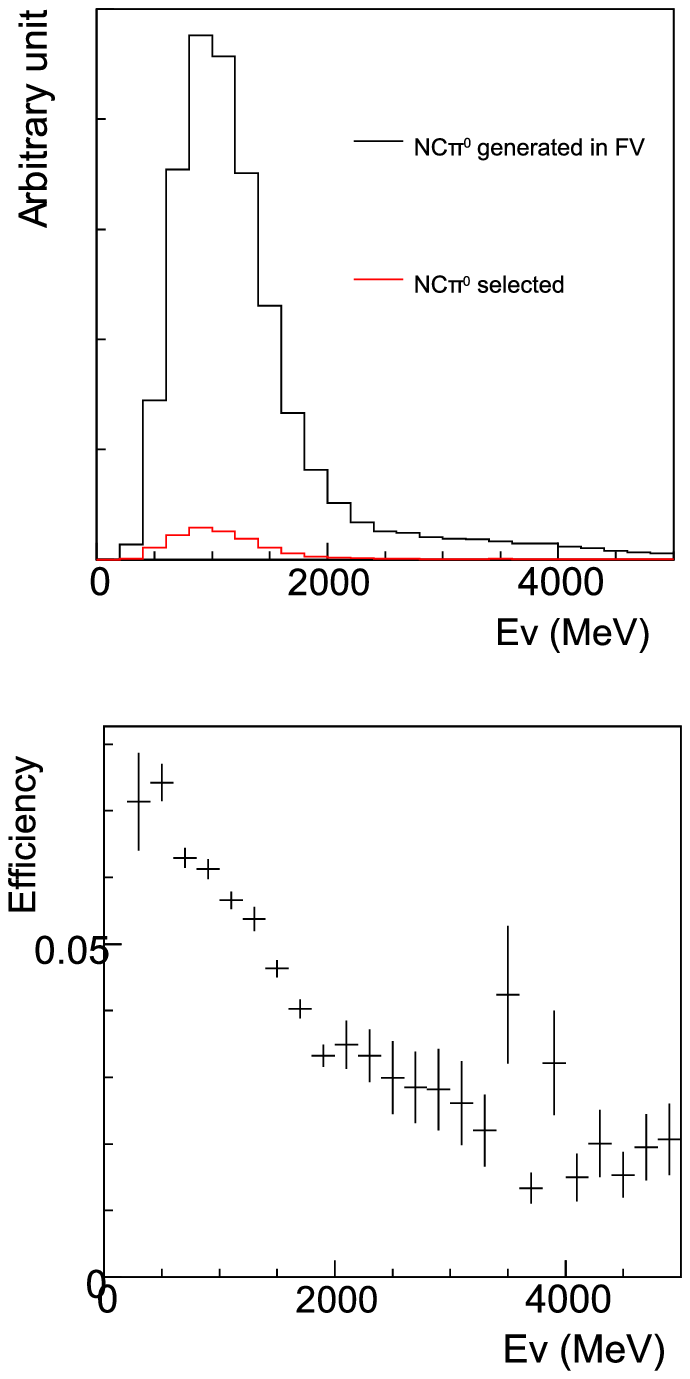}
      \caption{Expected neutrino energy spectra and selection efficiency as a function of 
	neutrino energy for NC$\pi^{\rm{0}}$ events.}
      \label{fig:enuncpi0}
  \end{center}
\end{figure}

The efficiency corrected number of NC$\pi^{\rm{0}}$ events is calculated 
as:
\begin{eqnarray}
N({\rm{NC}}\pi^{\rm{0}}) = \frac{N_{\rm{obs}}-N_{\rm{BG}}}{{\epsilon}_{{\rm{NC}}\pi^{\rm{0}}}},
\end{eqnarray}
where $N_{\rm{obs}}$ is the number of observed events, $N_{\rm{BG}}$ is the number of
background events estimated by the MC simulation, and ${\epsilon}_{\rm{NC}\pi^{\rm{0}}}$
is the selection efficiency of
NC $\pi^{\rm{0}}$ events calculated by the MC simulation.
$N_{\rm{obs}}$ and $N_{\rm{BG}}$,  ${\epsilon}_{\rm{NC}\pi^{\rm{0}}}$ are 657, 240.0 and 0.053, respectively. 
After subtracting background and correcting for the selection efficiency,
the number of NC$\pi^{\rm{0}}$ candidates is measured to be $[7.8 \pm 0.5 ({\rm stat.})] \times 10^2$. 
For the background calculation, we use the MC expectation normalized to the number of MRD stopped 
events. 
The neutrino energy dependence of the selection
efficiency for NC$\pi^{\rm{0}}$ events is shown in
Fig.~\ref{fig:enuncpi0}. The mean neutrino energy for NC\pzero\ 
events in the sample is estimated to be 1.1 GeV after event selection cuts.

\subsubsection{Total CC Interactions}
\label{sssec:cc}
\begin{figure}[htbp]
  \begin{center}
    \includegraphics[keepaspectratio=false,width=70mm,trim=0mm 30mm 50mm 50mm,clip]{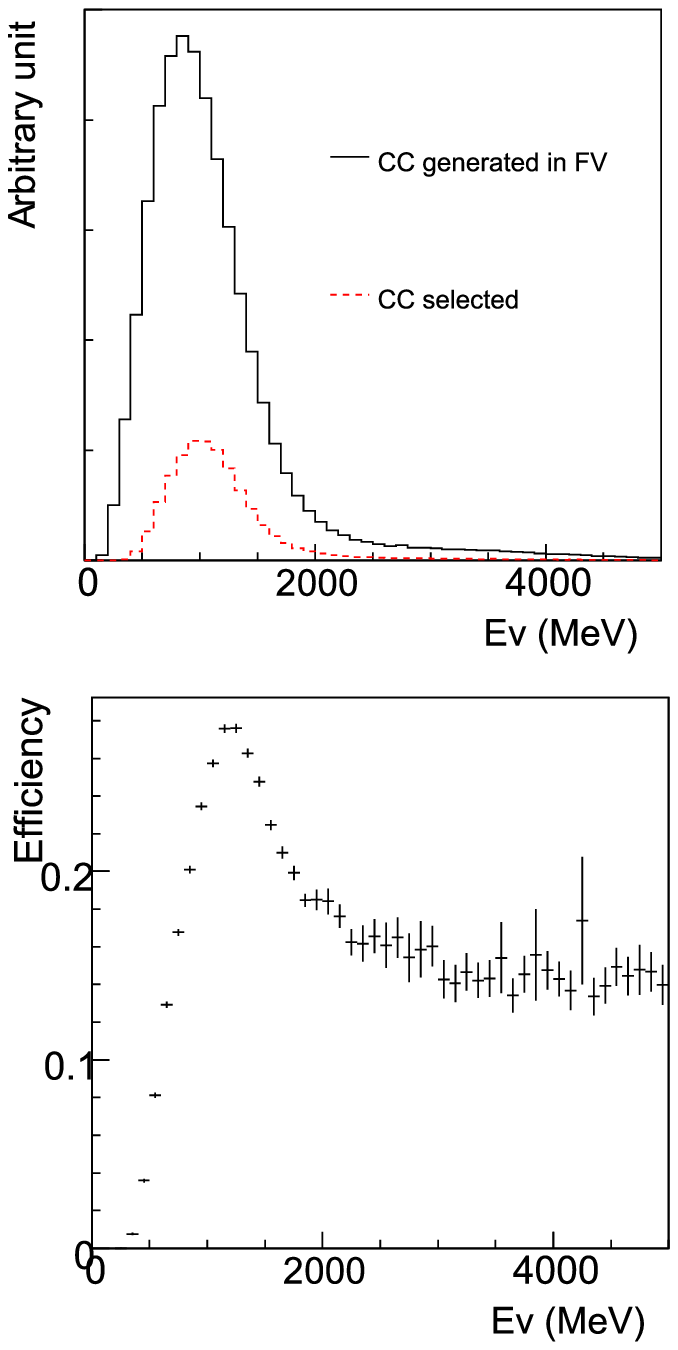}
    \caption{Expected neutrino energy spectra and selection efficiency as a function of 
      neutrino energy for all CC events.}
    \label{fig:enucc}
  \end{center}
\end{figure}

The total number of CC interactions is estimated using the MRD 
stopped sample. The mean neutrino energy of MRD-stopped events is estimated to be
1.2~GeV.
The number of CC candidates after 
correcting for the selection efficiency is calculated as:
\begin{eqnarray}
N({\rm CC}) = \frac{N^{\rm CC}_{\rm obs}-N^{\rm CC}_{\rm BG}}{{\epsilon}_{\rm CC}},
\end{eqnarray}
where $N^{\rm CC}_{\rm obs}$ is the number of observed CC events,
$N^{\rm CC}_{\rm BG}$ and $\epsilon_{\rm CC}$ are the number of background events and selection efficiency
in the sample, respectively, estimated with the MC simulation.
We observed 21,702 MRD-stopped events ($N^{CC}_{\rm{obs}}$). 
The number of background events and the selection efficiency 
are estimated to be 2348 ($N^{\rm CC}_{\rm BG}$) and 19\% ($\epsilon_{\rm CC}$),
respectively. The neutrino energy dependence of the selection efficiency for CC
 events is shown in Fig.~\ref{fig:enucc}. After subtracting the background events 
and correcting for the efficiency, the number of CC events is measured
to be $[1.02 \pm 0.01(stat.)] \times 10^{5}$.

\subsubsection{Cross section ratio}
\label{sub:sigratio}
The ratio of the NC \pzero\ production to the total
CC cross section is measured to be:
\begin{eqnarray}
\frac{\sigma({\rm{NC}}\pi^{\rm{0}})}{\sigma({\rm{CC}})}
&=& \frac{N({\rm{NC}}\pi^{\rm{0}})}{N({\rm CC})} \nonumber \\
&=& (7.7 \pm 0.5({\rm stat.}) \pm 0.5({\rm sys.})) \times 10^{-2},
\end{eqnarray}
at the mean neutrino energy of 1.14 GeV; systematic uncertainties are described in section~\ref{subsec:systematics}.
The MC expectation based on the Rein and Sehgal model is
$6.8 \times 10^{-2}$. The total uncertainty, 
$\pm0.7\times 10^{-2}$ is obtained adding statistical 
and systematic uncertainties in quadrature.
Although the value of this measurement
is larger than the expectation by 11\%, the excess 
corresponds to 1.3 standard deviations if the total
uncertainty is taken into account. 

\subsubsection{Systematic errors}
\label{subsec:systematics}
The sources of systematic error are divided into four categories,
 (i) detector response and track reconstruction, (ii) nuclear effects
 and neutrino interaction models, 
(iii) neutrino beam and (iv) dirt background. We vary these 
sources within their uncertainties and take the resulting change in the cross 
section ratio as the systematic uncertainty of the measurement. 
Table~\ref{tab:systot} summarizes the systematic errors in the NC\pzero\ cross section ratio. The total systematic error 
is $\pm 0.5 \times 10^{-2}$ on the cross section ratio.

\begin{table}[htbp]
\caption{Summary of the systematic errors in the NC\pzero\ cross section
ratio.}
\begin{center}
\begin{tabular}{ccc}
\hline
\hline
Source & \multicolumn{2}{c}{error ($\times 10^{-2}$)}\\ \hline
Detector response & -0.39 & 0.38 \\
$\nu$ interaction & -0.25 & 0.30 \\
Dirt background & -0.10 & 0.10 \\ 
$\nu$ beam & -0.11 & 0.22 \\ \hline
Total & -0.48 & 0.54 \\
\hline
\hline
\end{tabular} 
\end{center}
\label{tab:systot}
\end{table}

\paragraph{Detector response and track reconstruction}

The crosstalk of the MA-PMT was measured to be 3.15$\pm$0.4\% for adjacent
channels and is varied within the measurement error. The single photoelectron
resolution of the MA-PMT is set to 50\% in the simulation, to
reproduce the observed $dE/dx$ distribution of cosmic muons. 
The absolute error is estimated to be $\pm$20\%.  
Hence, we vary the single photoelectron resolution by $\pm$20\%.
Birk's constant for
the SciBar scintillator was measured to be $0.0208 \pm 0.0023 \ {\rm
cm/MeV}$ \cite{Hasegawa:2006am} and is varied within the
measurement error. 
The hit threshold for track reconstruction is varied by $\pm$20\%. 
A 10\% difference of the total pion-carbon cross section 
is seen for higher energy pions between the GEANT4 simulation
and external measurements. Hence,
we vary the cross section by $\pm10\%$. The uncertainty of
the energy scale of gamma rays is estimated to be $\pm$3\%. 
We vary the reconstructed energy of extended tracks by $\pm$3\%.
For the uncertainty on reconstruction of the gamma direction, we study how the
difference between data and MC distributions changes when we change the gamma
direction reconstruction algorithm. We compare our standard algorithm with gamma
direction reconstruction obtained using extended tracks - resulting in poorer
angular resolution. We take this change as the uncertainty.
The largest contribution to the uncertainty in the cross section 
ratio  are the crosstalk of the MA-PMT (-0.00,+0.27) and the hit
threshold (-0.25,+0.05). 

\paragraph{Neutrino interaction models and nuclear effects}

The uncertainty in CC resonant pion production is estimated to be
approximately $\pm 20$\% based on the K2K measurement
\cite{Rodriguez:2008eaa}. We vary the cross section of CC resonant
pion production by $\pm 20$\% and take that change as the systematic
error. We also vary the NC/CC ratio by $\pm 20$\% and take that
change as a systematic error. The uncertainty in the axial vector mass
for CC quasi-elastic and NC elastic scattering as well as CC(NC)
resonant pion production is estimated to be approximately $\pm
0.1$~GeV/$c^2$ based on recent measurements
\cite{Gran:2006jn,:2007ru}; results from past experiments are
systematically lower than recent measurements \cite{Bernard:2001rs},
and thus we only vary $M_A$ down to 1.11 GeV/$c^2$, and take that
change as the systematic error. The biggest contribution to the
uncertainty of the cross section ratio is the CC resonant pion
production (-0.14,+0.16).

We consider uncertainties in the pion interactions inside the
nucleus.  For pions produced by neutrino interactions,
uncertainties on the cross sections for pion absorption, pion
inelastic scattering and pion charge exchange in the nucleus are approximately
30\% \cite{Ashery:1981tq} in the momentum range of pions from $\Delta$ decays; 
we vary these pion interaction cross sections and take the resultant 
change in the ratio as the uncertainty. The largest
contribution to the uncertainty of the cross section ratio is 
the pion absorption (-0.17,+0.19). 

As a cross check, we measure the cross section ratio 
using the NUANCE event generator~\cite{Casper:2002sd} to predict event rates
and calculate efficiencies, and obtain a measured ratio of $7.9 \times 10^{-2}$ 
(the NUANCE expectation is $7.1 \times 10^{-2}$). 
The result using NUANCE agrees with the NEUT result ($7.7 \times 10^{-2}$)
within the systematic uncertainty, so we do not 
add the NEUT/NUANCE difference to the systematic uncertainty.

\paragraph{Dirt Backgrounds}

As shown in Fig.~\ref{fig:vtx_z_merged}, the dirt background
simulation describes data at $z~<~-20~{\rm cm}$ where the dirt
background is the dominant contamination.  However, the statistical
uncertainty is large, 15\%.  We scale the dirt contamination by
$\pm$15\% in the final sample and take the change as the systematic
error due to dirt backgrounds.

\paragraph{Neutrino Beam}

The uncertainties in secondary particle production cross sections in
proton-beryllium interactions, hadronic interactions in the target or
horn, and the horn magnetic field model are varied within their
externally estimated error bands. Detailed descriptions of each
uncertainty are found elsewhere \cite{AguilarArevalo:2008yp}.
Systematic uncertainties in the neutrino flux are reduced by removing
the model dependent parameterization in the propagation of errors from
the HARP data\cite{AguilarArevalo:2008yp}. Uncertainties associated
with the delivery of the primary proton beam to the beryllium target
and the primary beam optics, which result in an overall normalization
uncertainty, are not considered in this analysis since they cancel in
the cross section ratio.


\subsection{Reconstructed \pzero\ Kinematics}
\label{subsec:pi0rec}
\begin{figure}[tbp]
  \begin{center}
    \includegraphics[keepaspectratio=false,width=70mm,trim=0mm 20mm 0mm 0mm,clip]{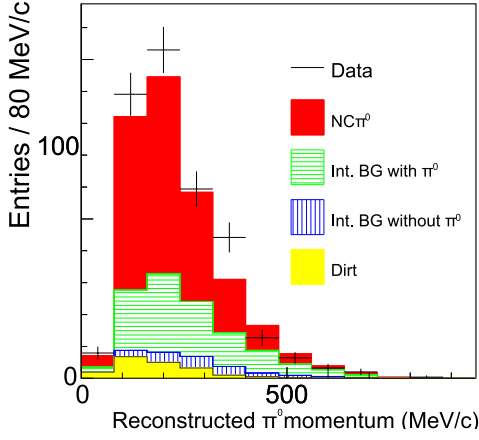}
  \end{center}
  \caption{The reconstructed \pzero\ momentum after all event selection cuts.}
  \label{fig:pi0mom_merged}
\end{figure}

\begin{figure}[tbp]
  \begin{center}
    \includegraphics[keepaspectratio=false,width=70mm,trim=0mm 10mm 0mm 0mm,clip]{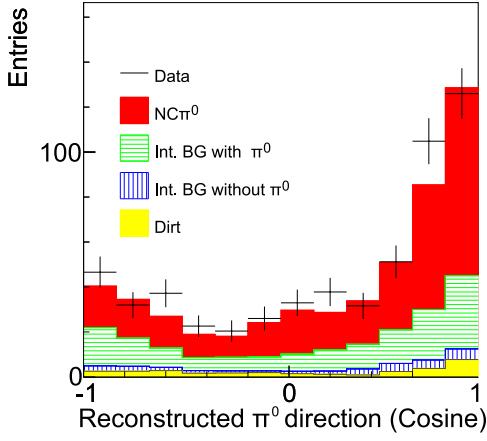}
  \end{center}
  \caption{Cosine of the reconstructed \pzero\ angle with respect to the beam direction after all event selection cuts.}
  \label{fig:pi0angle_cos_merged}
\end{figure}

\begin{figure}[tbp]
  \begin{center}
    \includegraphics[keepaspectratio=false,width=70mm,trim=0mm 30mm 40mm 40mm,clip]{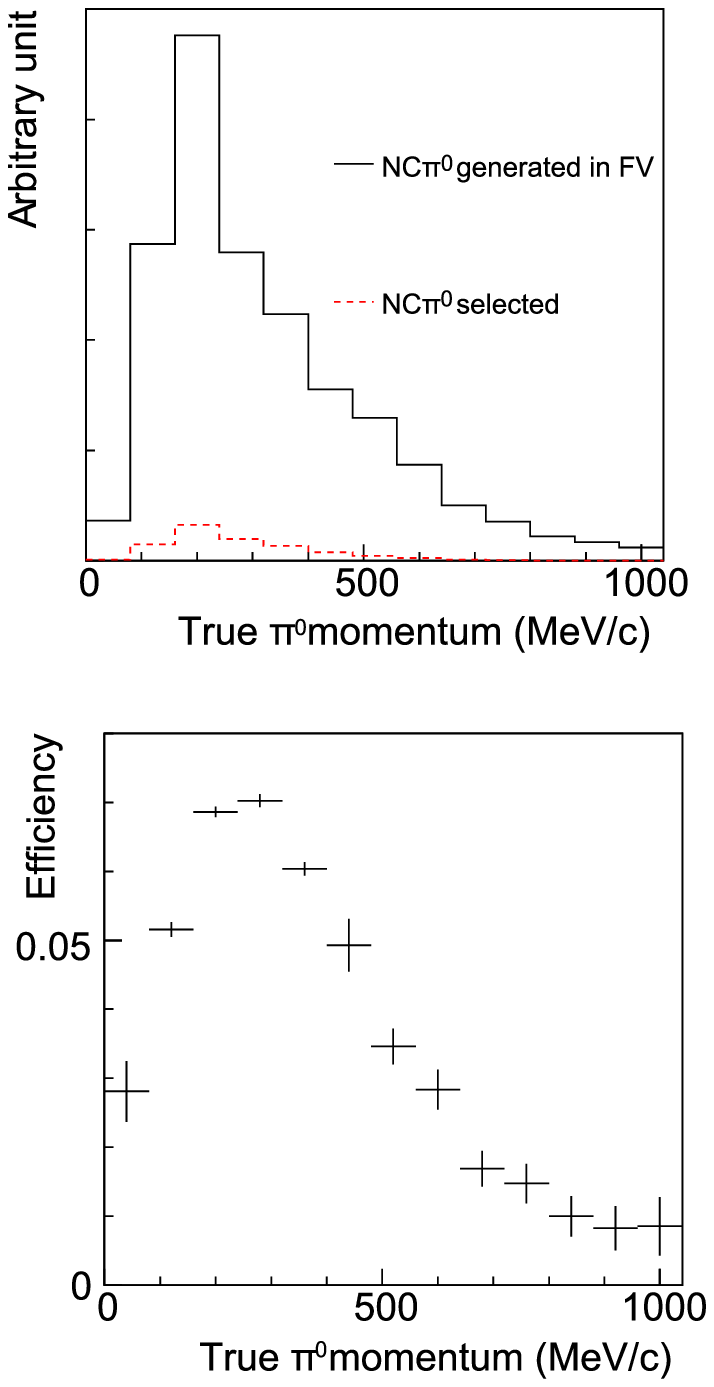}
  \end{center}
  \caption{The true \pzero\ momentum of
generated MC events (solid) and selected MC events (dashed) for all
NC\pzero\ production processes (top), and the true \pzero\ momentum
dependence of the efficiency of NC\pzero\ production (bottom).}
 \label{fig:pi0mom_efficiency}
\end{figure}

\begin{figure}[tbp]
  \begin{center}
    \includegraphics[keepaspectratio=false,width=70mm,trim=0mm 30mm 40mm 40mm,clip]{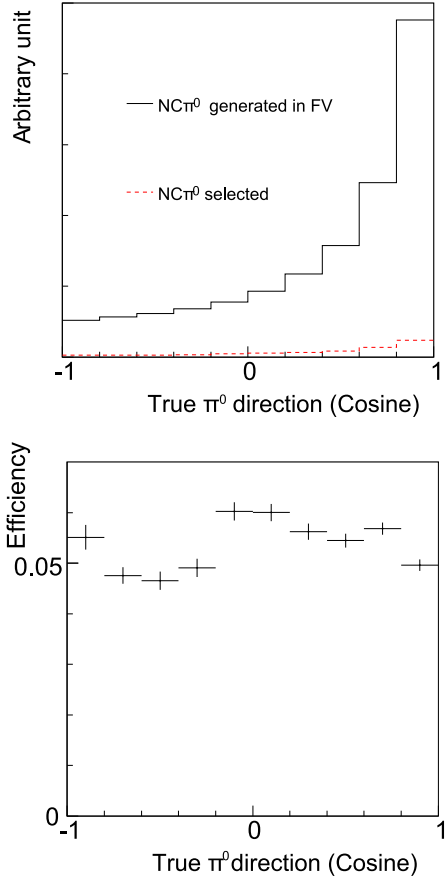}
  \end{center}
  \caption{The true \pzero\ angle with respect to the beam direction
for generated MC events (solid) and selected MC events (dashed) for
NC\pzero\ production (top), and the efficiency for NC \pzero\
production as a function of the true \pzero\ direction (bottom).}
 \label{fig:pi0angle_cos_efficiency}
\end{figure}

\begin{figure}[tbp]
  \begin{center}
    \includegraphics[keepaspectratio=false,height=70mm,width=70mm,trim=0mm 10mm 0mm 0mm,clip]{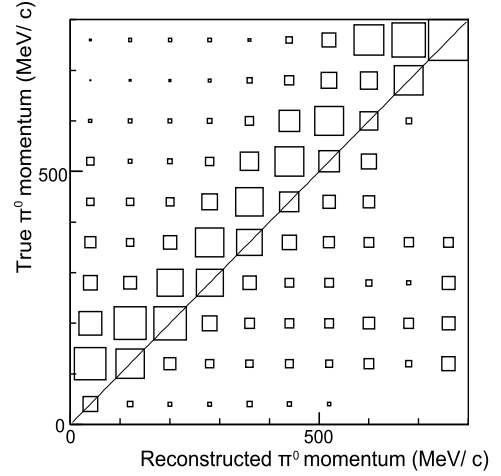}
  \end{center}
  \caption{The true \pzero\ momentum ($P^{\rm true}$) versus
 reconstructed \pzero\ momentum ($P^{\rm rec}$) from the MC simulation. 
 The solid line shows the identity. 
  }
  \label{fig:pi0mom_true2rec}
\end{figure}

\begin{figure}[tbp]
  \begin{center}
    \includegraphics[keepaspectratio=false,height=70mm,width=70mm,trim=0mm 0mm 0mm 0mm,clip]{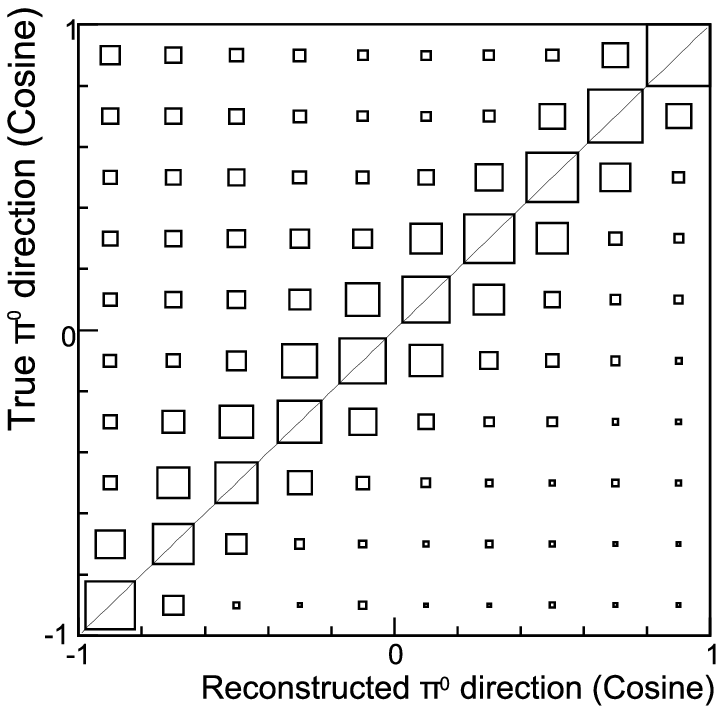}
  \end{center}
  \caption{The true \pzero\ direction ($\cos{\theta}^{\rm true}$)
    versus reconstructed \pzero\ direction  ($\cos{\theta}^{\rm rec}$)
    from the MC simulation.  The solid line shows the
    identity. 
  }
  \label{fig:pi0angle_cos_true2rec}
\end{figure}

After all event selection cuts, we studied the reconstructed kinematics of the 
\pzero s: the \pzero\ momentum and cosine of the
\pzero\ angle with respect to the beam direction, as shown in Fig.~\ref{fig:pi0mom_merged}
and Fig.~\ref{fig:pi0angle_cos_merged}.
The NC\pzero\ efficiency as functions of \pzero\ momentum and 
angle are shown in Fig.~\ref{fig:pi0mom_efficiency} and 
\ref{fig:pi0angle_cos_efficiency}, respectively.
The average momentum of reconstructed {\pzero}s is estimated to be 223 ${\rm MeV}/c$\ 
while the average momentum of true {\pzero}s after all event selection cuts is 264 ${\rm MeV}/c$
according to our MC simulation.
This difference comes from energy leakage of gamma rays.
The relation between the true and reconstructed \pzero\ momentum is shown in Fig.~\ref{fig:pi0mom_true2rec}. 
The momentum resolution is estimated to be 23\%.
The relation between the true and reconstructed \pzero\ direction is shown in Fig.~\ref{fig:pi0angle_cos_true2rec}. 
The angular resolution of {\pzero}s is estimated to be 6$^{\circ}$. 
In events with two {\pzero}s, we choose the \pzero\ with the largest 
momentum when comparing the true and reconstructed kinematic quantities.



\begin{figure}[tbp]
  \begin{center}
    \includegraphics[keepaspectratio=false,height=70mm,width=70mm,trim=0mm 0mm 0mm 0mm,clip]{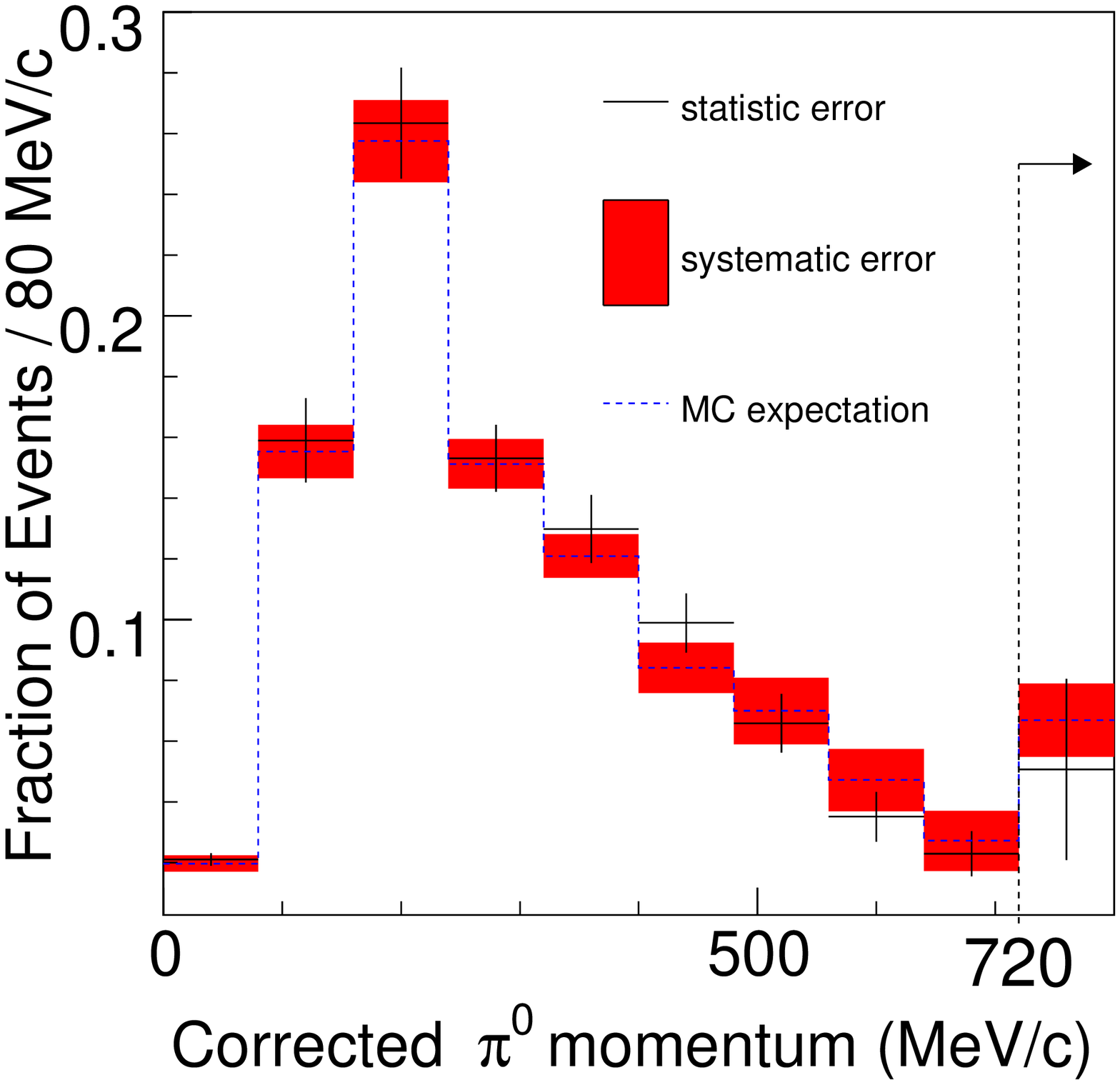}
  \end{center}
  \caption{The \pzero\ momentum distribution after all corrections
described in the text, with statistical (error bars) and systematic
(red boxes) uncertainties.  The dashed line shows the Monte Carlo
expectation based on the Rein and Sehgal model.}
  \label{fig:pi0mom_allcor}
\end{figure}

\begin{figure}[tbp]
  \begin{center}
    \includegraphics[keepaspectratio=false,height=70mm,width=70mm,trim=0mm 0mm 0mm 0mm,clip]{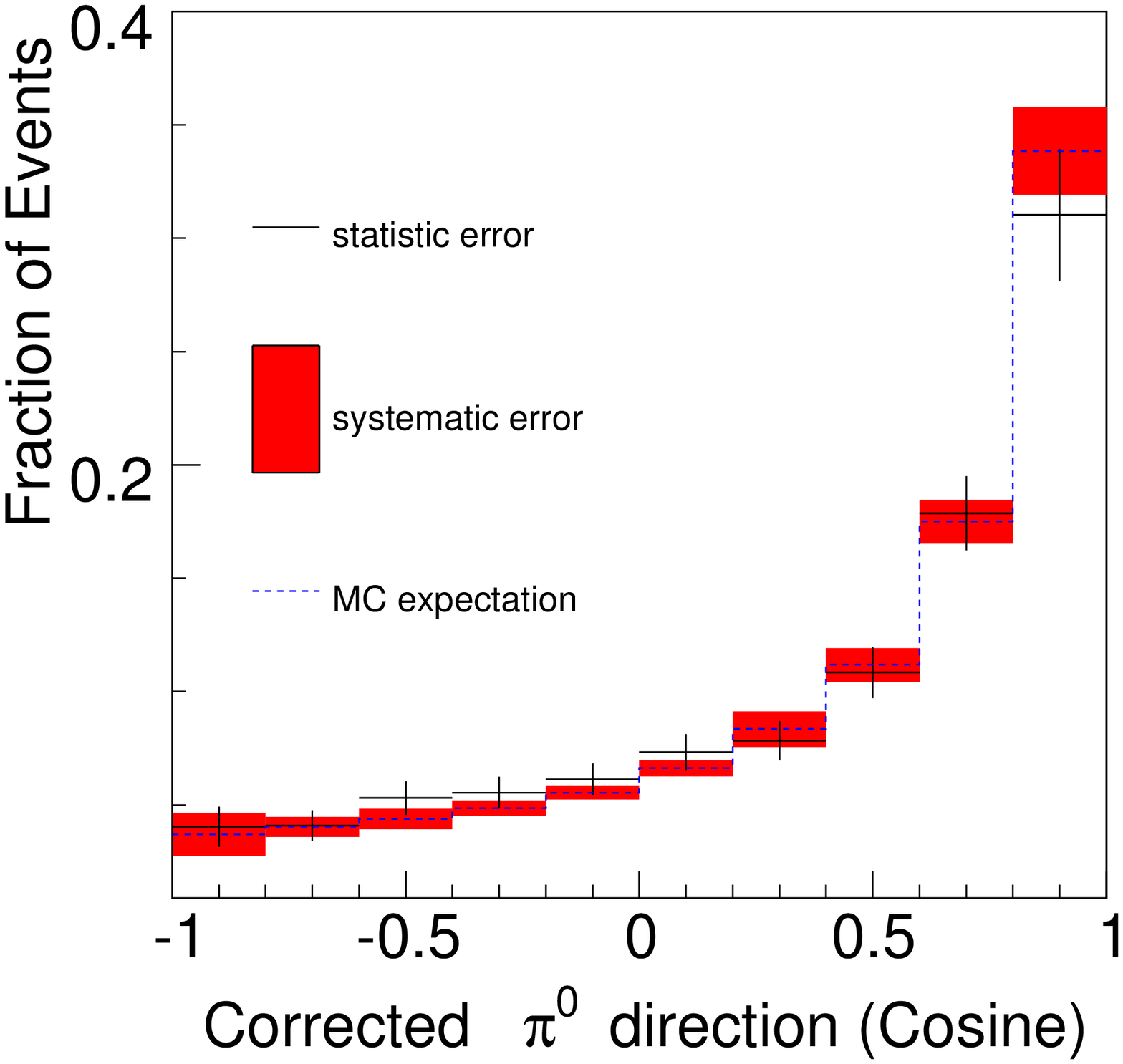}
  \end{center}
  \caption{The \pzero\ angular distribution after all corrections
described in the text, with statistical (error bars) and systematic
(red boxes) uncertainties.  The dashed line shows the Monte Carlo
expectation based on the Rein and Sehgal model.}
  \label{fig:pi0angle_cos_allcor}
\end{figure}

In the reconstructed \pzero\ momentum and angular distributions, 
we extract NC\pzero\ signal events by subtracting the expected backgrounds; 
to estimate backgrounds, we use the MC expectation normalized 
to the number of MRD-stopped events. 
After the background subtraction, we convert the reconstructed \pzero\ momentum (direction)
distribution to the true momentum (direction) distribution using a Bayesian unfolding
method~\cite{D'Agostini:1994zf}  using the MC simulation to define the unfolding matrix.  
Figure~\ref{fig:pi0mom_true2rec} (\ref{fig:pi0angle_cos_true2rec}) shows the true versus
reconstructed \pzero\ momentum (angle) distributions; these figures are graphical representations
of the smearing matrices used for the unfolding.
Finally, we perform the efficiency correction to obtain
the true \pzero\ momentum (direction) distribution.
Figures~\ref{fig:pi0mom_allcor} and \ref{fig:pi0angle_cos_allcor} show 
the \pzero\ momentum and direction distributions, respectively, 
after background subtractions, conversions to the true \pzero\ kinematics 
and efficiency corrections.
To compare the shapes of the distributions, the total numbers
of entries in the distributions are normalized to unity both
for the  measurement and the MC expectation.
The shapes of these two distributions agree with the MC expectation. 
The systematic errors of Fig.~\ref{fig:pi0mom_allcor}
and \ref{fig:pi0angle_cos_allcor} are expected to arise from the same sources, 
and are estimated in the same manner, as described in Sec.~\ref{subsec:systematics}.



\subsection{Coherent pion production}
\begin{figure}[htbp]
  \begin{center}
    \includegraphics[keepaspectratio=false,width=70mm,trim=0mm 0mm 0mm 0mm,clip]{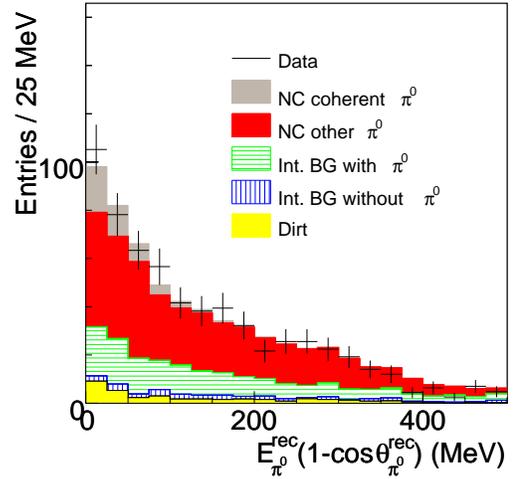}
 \end{center}
 \caption{ $E^{\rm rec}_{{\pi}^{0}}(1-\cos{\theta}^{\rm rec})$ after fitting. The coherent contribution and other NC\pzero\ are separately shown for the MC simulation.}
 \label{fig:pi0cohangleafterfit}
\end{figure}

In the coherent pion production, the neutrino interacts
the entire nucleus. 
In this case, the following relation should be satisfied. 
\begin{eqnarray}
\label{eq:cohshiki}
|t|  <  \frac{1}{R},  
\end{eqnarray}
where $t$ and $R$ are
the momentum transfer to the nucleus and 
the nuclear radius, respectively. 
Using Eq.~\ref{eq:cohshiki}, we can deduce 
\begin{eqnarray}
\label{eq:cohshiki2}
 E_{\pi^{0}}(1-\cos\theta_{\pi^{0}}) < \frac{1}{R} \sim {\rm ~100~MeV}  
\end{eqnarray}
according to Ref.~\cite{Lackner:1979ax}. In this equation, 
the $E_{\pi^{0}}$ and $\theta_{\pi^{0}}$ are the \pzero\ energy 
and direction with respect to the neutrino beam, respectively. 
Hence, the fraction of coherent \pzero\ production is extracted
from the $E^{\rm rec}_{{\pi}^{0}}(1-\cos{\theta}^{\rm rec})$ 
distribution shown in Fig.~\ref{fig:pi0cohangleafterfit}, 
where $E^{\rm rec}_{{\pi}^{0}}$ is the reconstructed \pzero\
energy calculated as $E^{\rm rec}_{{\gamma}1} + E^{\rm rec}_{{\gamma}2}$.  
We fit this distribution using three templates made by dividing 
the final MC sample into NC coherent \pzero, NC resonant \pzero   
and background samples.
Two parameters, ${\rm R}_{\rm coh}$, ${\rm R}_{\rm res}$, scale
NC coherent \pzero and NC resonant \pzero templates independently. 
The scale of the background sample is fixed to unity. The expected 
number of events in the $i$-th ($i=1,2,\ldots ,N(=20)$) bin in the  $E^{\rm rec}_{{\pi}^{0}}(1-\cos{\theta}^{\rm rec})$ 
distribution is expressed as: 
\begin{eqnarray}
  N^{\rm exp}_{i} = {\rm R}_{\rm coh}\times N^{\rm coh}_{i} 
+  {\rm R}_{\rm res}\times N^{\rm res}_{i}
+  N^{\rm BG}_{i}
\end{eqnarray}  
The fit minimizes the following ${\chi}^2$:
\begin{eqnarray}
 {\chi}^2 &=& -2\ln{\frac{f(N^{\rm obs};N^{\rm exp})}{f(N^{\rm obs};N^{\rm obs})}},
\end{eqnarray}
where $N^{\rm obs(exp)}$ represents the observed (expected) number of events in all bins
$(N^{\rm obs(exp)}_{1},N^{\rm obs(exp)}_{2},{\ldots},N^{\rm obs(exp)}_{N})$ 
and $f(N^{\rm obs};N^{\rm exp})$ is the Poisson likelihood to find $N^{\rm obs}$ events
when $N^{\rm exp}$ events are expected.
When the systematic errors for each bin and their correlation expressed using the covariance
 matrix $V_{jk}$ ($j,k=1,2,\ldots ,N(=20)$) are given, the likelihood is expressed as
\begin{eqnarray}
&& f(N^{\rm obs};N^{\rm exp};V)
= A\int_{}^{} \biggl[\Bigl[\prod^{N}_{i=1} 
dx_{i}\frac{{x_{i}}^{N^{\rm obs}_{i}}e^{-{x_{i}}}}{N^{\rm obs}_{i}!}\Bigr]\nonumber \\
&& \times \exp\Bigl[-\frac{1}{2}\sum_{j=1}^{N}\sum_{k=1}^{N}(x_{j}-N^{\rm exp}_{j})V^{-1}_{jk}(x_{k}-N^{\rm exp}_{k})\Bigr]\biggr],
\end{eqnarray} 
where $A$ is the normalizing constant. 
To calculate this integral, we generate 1000 MC expectations with random variations 
drawn from Gaussian distributions about the expectations for each bin, with correlations, 
estimated from the MC simulation.  Using $x_{i,m}$ for the $m$-th expectation in the $i$-th bin, 
the likelihood is expressed as :
\begin{eqnarray} 
 f(N^{\rm obs};N^{\rm exp};V) = \frac{1}{M}\sum_{m}\prod_{i}\frac{x^{N^{\rm obs}_{i}}_{i,m}e^{-x_{i,m}}}{N^{\rm obs}_{i}},
\end{eqnarray}
where $M$ is the total number of random samples (1000).  
The result of the fit is:
\begin{eqnarray}
{\rm R}_{\rm coh} &=& 0.56 \pm 0.34, \\
{\rm R}_{\rm res} &=& 1.33 \pm 0.16 .
\end{eqnarray}  
The  $E^{\rm rec}_{{\pi}^{0}}(1-\cos{\theta}^{\rm rec})$ distribution after the fitting is shown
in Fig.~\ref{fig:pi0cohangleafterfit}. The $\chi^2$ per degree of freedom (DOF), before the fit
is 17.8/20 = 0.89, and it is 12.9/18 = 0.72 after the fit. The statistical 
error and all systematic errors described in Sec.~\ref{subsec:systematics} are included in the errors of ${\rm R}_{\rm coh}$ and ${\rm R}_{\rm res}$. 
Without the systematic errors, we obtain $0.79 \pm 0.30({\rm stat.})$ and $1.24 \pm 0.13({\rm stat.})$
for ${\rm R}_{\rm coh}$ and ${\rm R}_{\rm res}$, respectively.  
The dominant systematic source is the detector response.  

The ratio of the NC coherent \pzero\ production
to the total CC cross sections from the MC prediction based on the Rein and
Sehgal model is $1.21~\times10^{-2}$. Hence, the cross section ratios 
are measured to be:
\begin{eqnarray}
\frac{\sigma({\rm NC coh}\pi^{\rm{0}})}{\sigma({\rm{CC}})}
&=& {\rm R}_{\rm coh} \times 1.21~\times10^{-2},  \nonumber \\
&=& (0.68 \pm 0.41) \times 10^{-2},
\end{eqnarray}
where ${\rm R}_{\rm coh}$ is 0.56$\pm$0.34. The mean neutrino energy for NC coherent
\pzero\ events in the sample is estimated to be 1.0 GeV.
This result is 1.6 standard deviations above the no coherent production assumption and
consistent with the MC prediction based on the Rein and Sehgal model.

\section{Conclusions}
\label{sec:conclusions}

In conclusion, we have observed the production of the NC\pzero\ events
by a muon neutrino beam on a polystyrene target ($\rm C_{8}H_{8}$)
using the SciBooNE neutrino data set of $0.99\times 10^{20}$ protons
on target. The ratio of the NC\pzero\ production to total CC cross
sections is measured to be $(7.7 \pm 0.5({\rm stat.}) \pm 0.5 ({\rm
sys.})) \times 10^{-2}$ at mean neutrino energy 1.1 GeV. The MC
prediction based on the Rein and Sehgal model~\cite{Rein:1980wg} is
$6.8 \times 10^{-2}$. The measured shapes of the \pzero\ momentum and
angular distributions, as shown in Figures~\ref{fig:pi0mom_allcor} and
\ref{fig:pi0angle_cos_allcor} agree with the MC prediction within uncertainties.
The ratio of NC coherent \pzero production to the total CC cross
section is measured to be $(0.7~\pm~0.4~)\times10^{-2}$ based on the
Rein and Sehgal model~\cite{Rein:1982pf}, while the MC
prediction is $1.21~\times10^{-2}$.

\section{Acknowledgements}
\label{sec:acknowledgments}

We acknowledge the Physics Department at Chonnam National University,
Dongshin University, and Seoul National University for the loan of
parts used in SciBar and the help in the assembly of SciBar.
We wish to thank the Physics Departments at
the University of Rochester and Kansas State University for the loan
of Hamamatsu PMTs used in the MRD.  We gratefully acknowledge support
from Fermilab as well as various grants, contracts and fellowships
from the MEXT and JSPS (Japan), the INFN (Italy), the Ministry of Science
and Innovation and CSIC (Spain), the STFC (UK), and the DOE and NSF (USA).
This work was supported by MEXT and JSPS with the Grant-in-Aid
for Scientific Research A 19204026, Young Scientists S 20674004,
Young Scientists B 18740145, Scientific Research on Priority Areas
``New Developments of Flavor Physics'', and the global COE program
``The Next Generation of Physics, Spun from Universality and Emergence''.
The project was supported by the Japan/U.S. Cooperation Program in the field
of High Energy Physics and by JSPS and NSF under the Japan-U.S. Cooperative
Science Program.

\end{document}